\documentclass[twocolumn, prd]{revtex4-2}
\usepackage[utf8]{inputenc}
\usepackage{amsmath}
\usepackage{amssymb}
\usepackage{geometry}
\usepackage{qcircuit}
\usepackage{import}\usepackage[utf8]{inputenc}
\usepackage{graphicx}
\usepackage{todonotes}
\usepackage{natbib}
\usepackage{color}
\usepackage{microtype}
\usepackage{import}
\usepackage{bbold}
\usepackage[plain]{fancyref}
\usepackage{varioref}
\usepackage{slashed}
\usepackage{tikz}
\usepackage{xcolor}
\usepackage[colorlinks]{hyperref}
\usepackage{cleveref}
\definecolor{winered}{rgb}{0.8,0,0}
\definecolor{darkb}{rgb}{0,0,0.8}
\hypersetup{
    colorlinks=true,
    citecolor=blue,
    linkcolor=winered,
    filecolor=red,      
    urlcolor=darkb,
}
\usetikzlibrary{shapes}

\begin{document}
\title{Prospects for Simulating a Qudit-Based Model of (1+1)d Scalar QED}
\author{Erik J. Gustafson}
\affiliation{University of Iowa, Department of Physics and Astronomy, Iowa City IA, 52242}
\date{April 2021}

\begin{abstract}
    We present a gauge invariant digitization of $(1+1)$d scalar quantum electrodynamics for an arbitrary spin truncation for qudit-based quantum computers. We provide a construction of the Trotter operator in terms of a universal qudit-gate set. The cost savings of using a qutrit based spin-1 encoding versus a qubit encoding are illustrated. We show that a simple initial state could be simulated on current qutrit based hardware using noisy simulations for two different native gate set.
\end{abstract}
\maketitle

\section{Introduction}
\label{sec:introduction}

Quantum computing offers a natural way to simulate the dynamics of quantum field theories. While classical Monte Carlo simulations of lattice gauge theories have been able to extract static quantities to high precision \cite{davoudi2020,padmanath2019hadron}, classical Monte Carlo simulations encounter problems with determinations of dynamic quantities due to a sampling to noise issue known as the sign problem. While work has been done to begin tackling these problems \cite{Alexandru:2016gsd, Kanwar:2021tkd} using classical computers, quantum computing still offers another path forward.

Many quantum field theories (QFT) have continuous symmetries such as Quantum Electrodynamics (QED) which has a $U(1)$ symmetry; Quantum Chromodynamics (QCD), which has a $SU(3)$ symmetry, and Scalar $\phi^4$ which has continuous values for the field. 
While classical computers can truncate these continuous symmetries to machine precision, in order to store the values of the nine matrix elements for a gauge link in QCD to double-precision would require $\mathcal{O}(1000)$ qubits. 
This is clearly infeasible for noisy intermediate scale quantum (NISQ) hardware and in practice truncations of these symmetries will be necessary.
These truncations of field or group elements can take various forms. The scalar fields in $\phi^4$ can be approximated with even distributions of the field values and imposing field cutoffs \cite{Jordan:2011ci, Klco_2019}. Compact QED can be mapped to a $\mathbb{Z}_n$ or $U(1)$ symmetries \cite{Bazavov_2015,bazavov2015effective,PhysRevLett.121.223201,Unmuth_Yockey_2018, Muschik_2017,Shaw_2020, Kaplan_2020,Raychowdhury_2020}.
The non-Abelian groups $SU(2)$ and $SU(3)$ can be digitized in various ways \cite{Alexandru_2019, Hackett_2019,Ji_2020, ciavarella2021trailhead, klco_2020}. Other methods such as imbedding the theory into higher dimensions using quantum link models \cite{Chandrasekharan:1996ih,Brower:1997ha} and D-Theory \cite{Beard_1998,BROWER2004149} is also possible.
Since truncations lead to a different theory being simulated on the computer, understanding how these truncations distort the physics is an interesting question but regardless these distortions must be removed \cite{Zohar_2013}.
Understanding how to return to the continuous symmetries is its own problem \cite{Alexandru_2019,2020arXiv200614160H,Hasenfratz:2001iz,Caracciolo_2001,PhysRevE.57.111,PhysRevE.94.022134,CARACCIOLO2001223, Raychowdhury_2020}. Digitizations of $U(1)$ for Quantum Electrodynamics typically use Hilbert spaces that have an odd integer states per site or link \cite{Bazavov_2015, bazavov2015effective, PhysRevLett.121.223201, Unmuth_Yockey_2018, Unmuth_Yockey_2019, Kaplan_2020, Raychowdhury_2020}; similar issues will arise for $SU(3)$ \cite{Alexandru_2019,Ji_2020}. Because these digitizations do not nicely map onto Hilbert spaces of dimension $2^n$ there are states that will not be used and will complicate the circuit structure. 

While simulations of Quantum Chromodynamics are still many years off, digital quantum simulations of $1+1$ and $2+1$ dimensional field theories are already in progress \cite{PhysRevA.79.062314, Martinez:2016yna, PhysRevA.87.032341, PhysRevLett.121.170501,CerveraLierta2018exactisingmodel, Barends_2016, Klco:2018kyo, Verdel_2020, Yamamoto_2020, 2020arXiv200909551B, Notarnicola_2020, Brower:2020huh}.
 Simulations of the Transverse Ising model (TIM) \cite{CerveraLierta2018exactisingmodel,Lamm:2018siq,GustafsonIsing,gustafson2019real, Kim_2020,yeteraydeniz2021scattering,vovrosh2020confinement,Kandala:2017aa,Kandala_2019,Salath__2015,Labuhn_2016,2017Natur.551..601Z,PhysRevE.58.5355,2017Natur.551..579B} and some simpler gauge theories such as the Schwinger model have been a major focus of qubit based computers \cite{Alexandru_2019,Zohar_2013, Martinez:2016yna, Klco:2018kyo,brower2020lattice}. Compact scalar Quantum Electrodynamics (sQED) in $(1+1)$d has implementations proposed for optical lattices \cite{bazavov2015effective,PhysRevLett.121.223201}. This model is also called the Abelian Higgs model, however we will refer to it as sQED in this work. sQED is a natural first step for simulations on near term qudit NISQ computers because it is a $(1+1)$d gauge theory with a continuous symmetry that is coupled matter. In addition this theory's Hamiltonian can be written in an explicitly local gauge invariant way \cite{Bazavov_2015,bazavov2015effective,PhysRevLett.121.223201,Unmuth_Yockey_2018}. The fact that this is a gauge-matter theory and the algebra for its Hamiltonian can naturally be represented by qudits makes this model amenable to simulations on near term NISQ computers using qudit-based architectures. This is not the only model amenable to qudit-based machines;  O(N) spin models and spin-1 Ising models are also possible \cite{Choi_2017} as well as (2+1)d U(1) \cite{Unmuth_Yockey_2019, Bender_2020,Zohar_2012}.  Simulations of dynamics for sQED would be timely given the recent interest in algorithms \cite{di2012elementary,baker2020efficient, 2007PhRvA.75b2313R,2015NatSR.514671G,Napolitano_2021, Gedik_2015,PhysRevLett.123.070505,Gokhale_2019}, testing \cite{Lapkiewicz_2011,Yurtalan_2020,kononenko2020characterization,2010PhRvL.105v3601B} and development of qutrit based hardware  \cite{PhysRevLett.100.060504, Blok:2020may,Zhang_2019, Yurtalan_2020, Veps_l_inen_2016, morvan2020qutrit,2020arXiv200303307B,PhysRevX.5.021026,PhysRevX.10.021060}. 

This paper is laid out as follows. Sec. \ref{sec:Model} discusses the Abelian Higgs model and its Hamiltonian formulation. In Sec. \ref{sec:Systematics}, we discuss the systematic errors introduced by spin truncation.  Sec. \ref{sec:Encoding} discusses how the Hamiltonian can be digitized on qudit base hardware. We walk through the choice of observable, the methods of state preparation, and the simulation using a noise model of a qutrit based quantum computer in Sec. \ref{sec:Simulation}. Finally, Sec. \ref{sec:conclusions} highlights the results and a road map of future models of interest. 
\section{Model}
\label{sec:Model}
Following closely \cite{bazavov2015effective,Bazavov_2015,PhysRevLett.121.223201,Unmuth_Yockey_2018}, $1+1$-d compact Scalar QED with the magnitude of the scalar field frozen to unity has the Euclidean lattice action, with similar notation as \cite{Bazavov_2015,bazavov2015effective,PhysRevLett.121.223201,Unmuth_Yockey_2018} is used for consistency,
\begin{equation}
    \label{eq:abelianhiggs}
    \begin{split}
    \mathcal{S} &= \mathcal{S}_{gauge} + \mathcal{S}_{matter}\\
    \mathcal{S}_{gauge} &= -\frac{1}{g^2 a_s a_{\tau}} \sum_x \sum_{\nu < \mu}\text{ReTr}(U_{x,\mu\nu})\\
    \mathcal{S}_{matter} &= -\kappa_s \sum_{x}\Big( \phi^{\dagger}_x U_{x,s} \phi_{x+\hat{s}} + h.c.\Big)\\
    & -\kappa_\tau \sum_{x}\Big( \phi^{\dagger}_x U_{x,\tau} \phi_{x+\hat{\tau}} + h.c.\Big),
    \end{split}
\end{equation}
where $\kappa_s = R^2 a_\tau / a_s$, $\kappa_\tau = R^2 a_s / a_\tau$, and $R$ is the radial scalar field magnitude and is generally allowed to vary but will be fixed to one in this work. 
Compact representations of the gauge and matter fields are used:
\begin{equation}
\begin{split}
    U_{x,\mu\nu} &= U_{x,\mu}U_{x + \mu, \nu}U^{\dagger}_{x + \nu, \mu}U^{\dagger}_{x, \nu},\\
    U_{x, \mu} &= e^{-i a_\mu g A_{x, \mu}},~\text{and } \phi_x = e^{i \theta_x}.
\end{split}
\end{equation}

After taking the continuous time limit and using the same notation as in \cite{bazavov2015effective,Bazavov_2015,PhysRevLett.121.223201,Unmuth_Yockey_2018} we find the following Hamiltonian
\begin{equation}
    \label{eq:Hamiltonian}
    \begin{split}
    \hat{H} &= \frac{U}{2} \sum_{i=1}^{N_s} (\hat{L}^z_i)^2 + \frac{Y}{2}\sum_{i=1}^{N_s - 1}(\hat{L}_i^z - \hat{L}_{i + 1}^z)^2\\
    & + \frac{Y}{2}\big((\hat{L}^z_1)^2 + (\hat{L}^z_{N_s})^2\big) - X \sum_{i = 1}^{N_s} \hat{U}_i^x 
    \end{split}
\end{equation}
where,
\begin{equation}
\label{eq:operatoralgebra}
    \begin{split}
        \hat{L}^z|m\rangle& = m|m\rangle\\
        \hat{U}^x &= \frac{1}{2}(\hat{U}^+ + \hat{U}^-),\\
        \text{ and } \hat{U}^{\pm}|m\rangle&= |m\pm1\rangle.
    \end{split}
\end{equation}
The coefficients in Eq. (\ref{eq:Hamiltonian}) are related to the lattice spacing and gauge coupling $U = g^2 a_s$, $Y=1 /2 R^2 a_s$ , $X=2 R^2 / a_s$. In theory the operators $\hat{L}^z$ and $\hat{U}^x$ are infinite dimensional with the values of $m$ in Eq. (\ref{eq:operatoralgebra}) ranging from $-\infty$ to $+\infty$. In practice a cut off will be necessary for implementation on quantum hardware so that the Hilbert space is finite. In this case the spins $m = -n_{max}, ..., 0, ... n_{max}$. The following subsections will highlight the digitization procedure for implementing the Hamiltonian in Eq. (\ref{eq:Hamiltonian}) on qudit based hardware as well as indicating the difficulties of implementation on qubit based hardware.

\section{Systematic Errors}
\label{sec:Systematics}
An important aspect of truncations is examining the size of the truncation versus the lattice spacing. It should be unsurprising that truncation effects will become more significant the closer to the continuum we go.
In order to measure the effectiveness of these truncations we will use the following quantity 
\begin{equation}
    \label{eq:chi}
    \chi = \frac{1}{N_s} \sum_{i = 1}^{N_s} \sum_{j = 1}^{N_s} \langle \Omega |(\hat{L}^z_i - \hat{L}^z_j)^2 |\Omega\rangle,
\end{equation}
where $|\Omega\rangle$ is the ground state
This quantity measures how correlated the fields are at different sites. 
This measure is likely more accurate than comparing eigenvalues of the Hamiltonian because it probes off-diagonal elements and excited states with respect to the eigenbasis of the Hamiltonian.

\begin{figure}[t]
\vskip-0.5em
\includegraphics[width=0.48\textwidth]{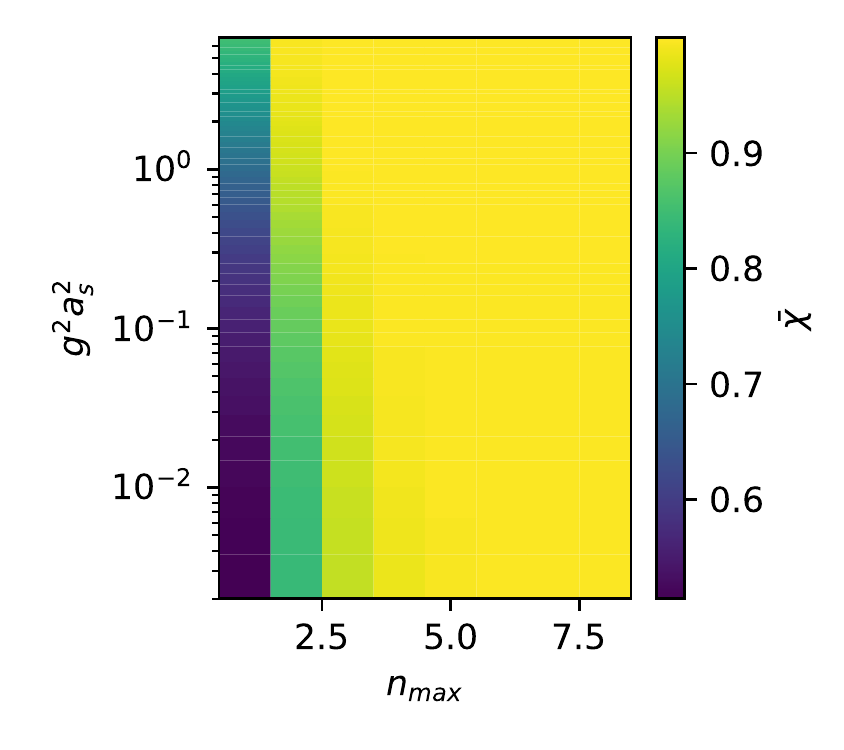}
\vskip-2em
\caption{$\chi$ normalized by an infinite bound extrapolation as a function of the coupling and spin truncation for 4 sites.}
\label{fig:truncations}
\vskip-2em
\end{figure}

We expect that at small coupling ($g^2 a_s^2$) since the high spin states are easily excited a larger truncation will be necessary. 
Conversely for strong coupling we should expect that a more coarse truncation will be acceptable. 
Fig. \ref{fig:truncations} shows $\chi$ normalized by the untruncated value as a function of the coupling, $g^2 a_s^2$, for a $n_s = 4$ site lattice. 
The supposition posited earlier carries out here.
For couplings on the order of $g^2 a_s^2 \leq 10^{-2}$, $n_{max} > 4 $ to effectively capture the physics desired. While for couplings of $g^2 a_s^2 \approx 1$, a truncation of $n_{max} = 2$ appears to be sufficient. 
One key feature that is evident is that there seems to be a stark difference between the spin-1 and spin-2 truncation at all couplings; this was seen as well in \cite{zhang2021truncation}. The discrepancy is not unexpected, for $Z_n$ theories there is a marked discrepancy between $n\leq 4$ and $n\geq5$ accurate representations of $U(1)$ \cite{PhysRevD.20.1915}. 

The key point that should be understood is that in the strongly coupled regime a spin-2 (qupet) will likely be sufficient but in the weakly coupled regime a spin-4 to spin-6 truncation will be necessary to capture the desired physics. 
In particular, the limit of $g^2 a_s \rightarrow 0$ Eq. (\ref{eq:Hamiltonian}) becomes that of the O(2) model in 1+1-dimensions. 
In this limit the couplings become $X Y=1$ in units of $a_s = 1$. At this ratio of couplings a $n_{max}=6$ spin truncation is effective to capture the desired physics \cite{Unmuth_Yockey_2018}.  
For nearer term devices a qupet may not yet be feasible but qutrits are actively being studies.
This will provide a good foundation for benchmarking and developing tools for higher truncations even if it does not accurately represent the physics of the theory.

\section{Encoding}
\label{sec:Encoding}

\subsection{qudit representation}
As previously mentioned, implementation on physical hardware requires a maximal spin cut off denoted $n_{max}$. For a given maximal integral spin, the operators defined in Eq. (\ref{eq:operatoralgebra}) behave as follows,
\begin{equation}
    (\hat{L}^z)_{i, j} = (n - i) \delta_{i, j}
\end{equation}
where $0 \leq i, j < 2 n + 1$ and $U^x$ can either truncate at $m=\pm n_{max}$ or have highest and lowest states wrap around like a $\mathbb{Z}_{n_{max}}$ theory. 

\begin{figure*}[ht]
    \centering
    \begin{gather*}
    \Qcircuit @C=1em @R=1em {
    & \gate{e^{-i\delta t (U + 2Y)\hat{L}^z}} &  \multigate{1}{e^{-2i\delta t Y\hat{L}^z\otimes \hat{L}^z}} & \qw & \qw & \gate{e^{i \delta t X \hat{U}^x}} & \qw \\
    & \gate{e^{-i\delta t (U + 2Y)\hat{L}^z}} & \ghost{e^{-2i\delta tY\hat{L}^z\otimes \hat{L}^z}} &  \multigate{1}{e^{-2i\delta t Y\hat{L}^z\otimes \hat{L}^z}} & \qw \qw & \gate{e^{i \delta t X \hat{U}^x}} & \qw \\
    & \gate{e^{-i\delta t (U + 2Y)\hat{L}^z}} &  \multigate{1}{e^{-2i\delta t Y\hat{L}^z\otimes \hat{L}^z}} &  \ghost{e^{-2i\delta tY\hat{L}^z\otimes \hat{L}^z}} & \qw \qw& \gate{e^{i \delta t X \hat{U}^x}} & \qw \\
    & \gate{e^{-i\delta t (U + 2Y)\hat{L}^z}} & \ghost{e^{-2i\delta tY\hat{L}^z\otimes \hat{L}^z}} & \qw & \qw & \gate{e^{i \delta t X \hat{U}^x}} & \qw
    }
    \end{gather*}
    \caption{Quantum circuit for $U_{tr}(\delta t)$ defined in Eq. (\ref{eq:trotter}).}
    \label{fig:trotter}
\end{figure*}
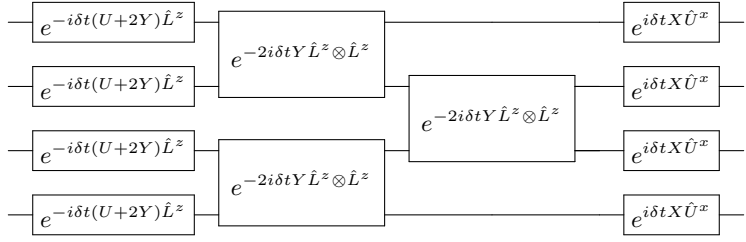

The Hamiltonian can be separated into two non-commuting parts.
The terms containing $L^z$ and the terms containing $U^x$. 
This spin-$n_{max}$ system is most naturally mapped onto a $(2n_{max}+1)$-qudit architecture. A universal basis for this machine will require a collection of $2n_{max} + 1$ $SU(2)$ rotations that couple the energies levels together and allow an arbitrary $SU(2n_{max} + 1)$ rotation. We can define these operators that compose the $SU(2)$ sub-algebras as generalizations of the Pauli matrices which will be defined as $\hat{X}^{a,b}$, $\hat{Y}^{a,b}$, and $\hat{Z}^{a,b}$. 
The behavior on these operators is given by
\begin{equation}
\hat{\sigma}^x_{a,b} |c\rangle = 
\begin{cases}
|b\rangle & c = a\\
|a\rangle & c = b\\
0 & c \neq a, b\\
\end{cases},
\end{equation}
\begin{equation}
\hat{\sigma}^y_{a,b} |c\rangle = 
\begin{cases}
-i|b\rangle & c = a\\
i|a\rangle & c = b\\
0 & c \neq a, b\\
\end{cases},
\end{equation}
and
\begin{equation}
\hat{\sigma}^z_{a,b} |c\rangle = 
\begin{cases}
|a\rangle & c = a\\
-|b\rangle & c = b\\
0 & c \neq a, b\\
\end{cases},
\end{equation}
where $1 \leq a,~b,~c \leq 2 n_{max} + 1$.
We can also define rotations,
\begin{equation}
\label{eq:quditrotations}
    \hat{R}^{\alpha}_{(a,b)}(\theta) = e^{i\theta \hat{\sigma}^{\alpha}_{(a,b)}},
\end{equation}
where $\alpha = x,y,z$, and $a$ and $b$ indicate the states for the Pauli sub-algebra to mix between.
For an arbitrary spin-$n_{max}$ system we can write the operators of  \Fref{eq:Hamiltonian} as follows:
\begin{equation}
    \label{eq:Lzsu2n1}
    \hat{L}^z = \sum_{j = 1}^{n_{max}} (n_{max}- j + 1)(\hat{\sigma}^z_{j, n_{max}} + \hat{\sigma}^z_{n_{max}, n_{max} + j})
\end{equation}
and
\begin{equation}
    \label{eq:Uxsu2n1}
    \hat{U}^x = \frac{1}{2}\Bigg(c_{bound}\hat{\sigma}^x_{(1, 2n_{max} + 1)} + \sum_{j = 1}^{2n_{max} - 1} \hat{\sigma}^x_{(j, j+1)}\Bigg)
\end{equation}
where $c_{bound}$ is $1$ if a $\mathbb{Z}_n$ model is desired or $0$ if a $U(1)$-truncation is desired. 

Time evolution of a state $|\psi\rangle$ is carried out via the traditional operator $e^{-i t \hat{H}}$. In order to implement this on a quantum computer we need to Trotterize \cite{Lloyd1073} the Hamiltonian and split it into non-commuting terms:
\begin{equation}
    \label{eq:trotter}
    \begin{split}
    \hat{U}_{tr}(\delta t) =& \Big(e^{-i \delta t (U  + 2 Y) / 2 \sum (\hat{L}^z_i)^2}\\
    &e^{-i \delta t Y \sum \hat{L}^z_{i}\hat{L}^z_{i + 1}} e^{i \delta t X \sum \hat{U}^x_i}\Big).
    \end{split}
\end{equation}
A diagram of this circuit for any spin truncation is shown in Fig. \ref{fig:trotter}. This Trotterization is straight forward; three types of terms will be present:
\begin{enumerate}
    \item one qudit rotation $e^{-i \delta t (U  + 2 Y) / 2 (\hat{L}^z)^2}$
    \item one qudit rotation $e^{i \delta t X \hat{U}^x}$
    \item two qudit rotations $e^{i \delta t Y L^z_i L^z_{i + 1}}$
\end{enumerate}
The $(\hat{L}^z)^2$ rotations are relatively straight forward to implement:
\begin{equation}
    \label{eq:Lz2rot}
    e^{i\theta (\hat{L}^z)^2} = \prod_{j = 1}^{2n_{max}} e^{i a_{j,j+1} \theta \hat{\sigma}^z_{(j, j+1)}}
\end{equation}
where the $a_{j,j+1}$ terms are found by solving the linear equation,
\begin{equation}
    (\hat{L}^z)^2 = \alpha_0 \mathbb{1} + \sum_{j=1}^{2n_{max} - 1} \alpha_{j,j+1} \hat{\sigma}^z_{j, j+1}
\end{equation}
for the coefficients $\alpha_{j, j+1}$ and $\alpha_0$. The values for these operators are found to scale quadratically with respect to the spin truncation $n_{max}$. With more details provided in Appendix \ref{app:a}.

The $\hat{L}^z_i \hat{L}^z_{i + 1}$ term is also relatively straight forward as well and involves solving a similar set of equations.
This operator can be written for arbitrary $n$ as,
\begin{equation}
    \label{eq:Lzlzrot}
    \begin{split}
    e^{-i\theta \hat{L}^z\hat{L}^z} =& C_{sum} \prod_{i=1}^{2n_{max} - 1} \\
    &\Bigg(\prod_{j = 1}^{2n_{max} - 1} R^{z;t}_{j, j+1}(\theta \beta_j)\Bigg) C_{sum}.
    \end{split}
\end{equation}
The $^t$ on single qutrit rotations indicates they are applied on the target qutrit of the $C_{sum}$ gates. 
This coupled $L^zL^z$ rotation then can be written in terms of at most $2n_{max} + 1$ two-qudit gates and $4n^2_{max}$ one qudit diagonal rotations which are expected to be relatively noiseless \cite{Blok:2020may}.
The $C_{sum}$ operator is a generalization of the CNOT gate and shifts the state $|a\rangle$ to $|a+1\rangle$; it is given by
\begin{equation}
\label{eq:csum}
    C_{sum} = \sum_{k = 1}^{2 n_{max} + 1} |k\rangle\langle k|\otimes \hat{\mathcal{X}}^k,
\end{equation}
where
\begin{equation}
\label{eq:genx}
\hat{\mathcal{X}} = \sum_{k=1}^{2n_{max}+1} |k\rangle \langle \text{mod}_{2n_{max} + 1} (k + 1)|.
\end{equation}

The $\hat{U}^x_i$ rotations can be implemented in one of two ways.
The naive way is to Trotterize the components of $\hat{U}^x$,
\begin{equation}
    e^{i\theta \hat{U}^x} \approx e^{i\theta c_{bound} \hat{\sigma}^x_{(1, 2n_{max} + 1)}}\prod_{j=1}^{2n_{max}} e^{i\theta \hat{\sigma}^x_{(j, j+1)}},
\end{equation}
which results in $2n_{max} + 1$ single qudit rotations. 
A slightly more complicated but exact way involves finding the set of rotations $\lbrace R^{\alpha}_{(a,b)}(\theta_j)\rbrace$ which implement $e^{i\theta \hat{U}^x}$.
This is relatively simple and tractable for most qudit based architectures because it involves diagonalizing a $2n_{max} + 1$ dimensional matrix where $n_{max}$ will be less than 12 for any approximation of $U(1)$. This will require at most $(2n_{max} + 1)^2 - 1$ rotations. In the case where $c_{bound} = 1$ this involves constructing a generalization of the Hadamard gate to a qudit \cite{Blok:2020may,morvan2020qutrit},

\begin{equation}
    \hat{\textbf{H}} = \sum_{k=0}^{2n}\sum_{j=0}^{2n} |k\rangle\langle j| e^{i(k + j)\pi/(2n + 1)}.
\end{equation}

\subsection{spin 1: Qubit vs. Qutrit}
Up until this point the work has been generalized to qudits. While high spin truncations are needed to simulate actual quantum electrodynamics \cite{Bazavov_2015,Unmuth_Yockey_2018}, qudits become more difficult to control the more states that are included \cite{Gokhale_2019,baker2020efficient}. From this point on we will specialize to the spin-1 truncation through out this work and set $c_{bound} = 0$. 
The fundamental operators from the Hamiltonian defined in Eq. (\ref{eq:Hamiltonian}) $\hat{L}^z$ and $\hat{U}^x$. For a spin-1 (3 state) truncation the $\hat{L}^z$ operator is defined as
\begin{equation}
    \hat{L}^z = \begin{pmatrix}
    1 & 0 & 0 \\
    0 & 0 & 0 \\
    0 & 0 & -1 \\
    \end{pmatrix}
\end{equation}
and can be embedded into a two qubit Hilbert space with the following encoding:
\begin{equation}
    \hat{L}^z = (\hat{Z}_2 + \hat{Z}_1 \otimes \hat{Z}_2) / 2,
\end{equation}
where the $\hat{Z}_{i}$ correspond to the Pauli-z matrix on qubit i. By extension, the operator $(\hat{L}^z)^2$ is given by
\begin{equation}
(\hat{L}^z)^2 = (1 + \hat{Z}_1) / 2.
\end{equation}
Similarly the $\hat{U}^x$ operator is given by 
\begin{equation}
\label{eq:uxu1}
    \hat{U}^x = \frac{1}{2}\begin{pmatrix}
    0 & 1 & 0 \\
    1 & 0 & 1 \\
    0 & 1 & 0 \\
    \end{pmatrix},
\end{equation}
and can be embedded into the Hilbert space of two qubits with the follwing linear combination of tensor products
\begin{equation}
    \begin{split}
        \hat{U}^x &= \hat{X}_1 \otimes (\mathbb{1} + \hat{X}_2 + \hat{Z}_2) / 2\\
        & + \hat{Y}_1 \otimes \hat{Y}_2 / 2.\\
    \end{split}
\end{equation}
The Trotterization can be broken up into the the two one-qutrit rotations,
\begin{equation}
    e^{-i\delta t (U/2 + Y)(L^z_i)^2} ~ \text{and} ~ e^{i \delta t X U^x},
\end{equation}
and the two-qutrit rotation,
\begin{equation}
    e^{i \delta t Y \hat{L}^z_i \hat{L}^z_{i + 1}}.
\end{equation}

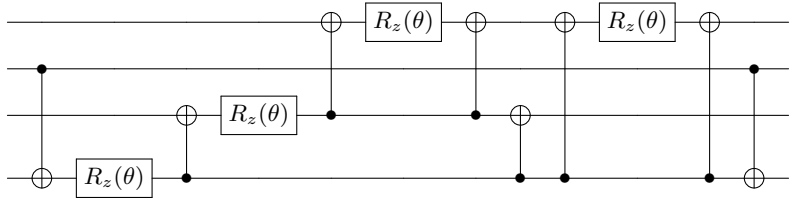
\begin{figure*}

    \begin{gather*}
    \Qcircuit @C=1em @R=1em {
    & \qw  & \qw & \qw & \qw & \targ & \gate{R_z(\theta)} & \targ & \qw & \targ & \gate{R_z(\theta)} & \targ & \qw & \qw \\
    & \ctrl{2} & \qw & \qw  & \qw & \qw & \qw & \qw & \qw & \qw & \qw & \qw & \ctrl{2} & \qw  \\ 
    & \qw & \qw & \targ & \gate{R_z(\theta)} & \ctrl{-2} & \qw & \ctrl{-2} & \targ & \qw & \qw & \qw & \qw & \qw \\
    & \targ & \gate{R_z(\theta)} & \ctrl{-1} & \qw & \qw & \qw & \qw & \ctrl{-1} & \ctrl{-3} & \qw & \ctrl{-3} & \targ & \qw\\
    }
    \end{gather*}
    \caption{circuit for $e^{i\theta\hat{L}^z\otimes\hat{L}^z}$ using a qubit embedding.}
    \label{fig:qubitlzlzrot}
\end{figure*}
In the case of physical qutrits, since both of the single qutrit rotations are an element of SU(3) they can easily be broken up into 8 rotations \cite{Tilma_2002} defined in Eq. (\ref{eq:quditrotations}) as follows:
\begin{equation}
\begin{split}
    \hat{V} =& e^{i \alpha_1 \hat{\sigma}_{01}^z} e^{i \alpha_2 \hat{\sigma}_{01}^y} e^{i \alpha_3 \hat{\sigma}_{01}^z}e^{i \alpha_4 \hat{\sigma}_{02}^y}\\& e^{i \alpha_5 \hat{\sigma}_{01}^z}e^{i \alpha_6 \hat{\sigma}_{01}^y} e^{i \alpha_7 \hat{\sigma}_{01}^z}e^{i \alpha_8 \hat{\sigma}_{12}^z},
    \end{split}
\end{equation}
where $\hat{V}$ is an arbitrary $SU(3)$ rotation.

Using this Euler decomposition we find that these operators can be written with the following rotations,
\begin{equation}
\label{eq:lz2rotdecomp}
\begin{split}
    e^{-\frac{i \delta t (U + 2Y)}{2}(\hat{L}^z)^2} =& R^z_{0,1}(-\delta t (U/6 + Y / 3))\\
    &R^z_{1,2}(\delta t (U / 6 + Y / 3)),
    \end{split}
\end{equation}
and 

\begin{equation}
\label{eq:uxrotdecomp}
\begin{split}
    e^{i \delta t X \hat{U}^x} =& \hat{R}^y_{0, 1}\Big(-\frac{\pi}{4}\Big) \hat{R}^y_{0, 2}\Big(\frac{\pi}{4}\Big)\hat{R}^z_{0, 1}\Big(\frac{\delta t X\sqrt{2}}{2}\Big)\\
    &\hat{R}^y_{0, 2}\Big(-\frac{\pi}{4}\Big)
    \hat{R}^y_{0, 1}\Big(\frac{\pi}{4}\Big).
    \end{split}
\end{equation}
Given the native gate set of \cite{Blok:2020may,morvan2020qutrit}, these two rotations together can be implemented in at most 15 one-qutrit rotations and likely fewer depending on the angles in the of the $\sigma^{ab}_y$ rotations; while the implementation of \cite{kononenko2020characterization} can implement this with at most 5 rotations for the $\hat{U}^x$ term and 2 for the $(L^z)^2$ term, the extra gates from \cite{Blok:2020may} come from the non-continuous parameterization of the $X$ and $Y$ rotations. Given that $\sigma_z^{ab}$ rotations are done virtually they are effectively noiseless and at most only 8 noisy gates are present. 

The two qutrit $L^z\otimes L^z$ rotation is implementable with 3 controlled sum gates,
\begin{equation}
    C_{sum} = \sum_{k=0}{2} |k\rangle\langle k|\otimes (X^{01}X^{12})^k,
\end{equation}
and 4 single qutrit $\sigma^z$ rotations. This implementation is,
\begin{equation}
\label{eq:lzlzrotdecomp}
\begin{split}
    e^{i \theta L^z\otimes L^z} &= C_{sum} R^{z;t}_{0,1}(2\theta / 3)R^{z;t}_{1, 2}(\theta / 3)\\
    &C_{sum} R^{z;t}_{1,2}(2\theta / 3)R^{z;t}_{0,1}(\theta / 3) C_{sum}.
    \end{split}
\end{equation}
The controlled sum is a generalization of the qubit-CNOT gate to qudit based architecture \cite{Di_2013} and is realizable on current qutrit based hardware \cite{morvan2020qutrit, Blok:2020may}. The $^t$ on the single qutrit rotations indicates that they are applied on the target qutrit of the $C_{sum}$ gate. 

In the context of the implementing on physical qubits, the $U^x$ operator will require 3 CNOTs to implement \cite{Vatan_2004}. The $\hat{L}^z\otimes \hat{L}^z$ term will require approximately 16 CNOTs to couple the all the 2-, 3-, and 4-qubit rotations (shown in Fig. \ref{fig:qubitlzlzrot}) 

At this point, it is clear the qutrit formulation clearly is better than the qubit formation because of the reduced number of entangling gates but a qubit formulation is possible. The circuit depth in the qutrit formulation is 6 two-qutrit gates deep per Trotter step, while the qubit formulation is 19 CNOTs deep per Trotter step. 

\begin{table}[!ht]
    \centering
    \begin{tabular}{c|cc|cc}
    \hline
         gate & \multicolumn{2}{c|}{qubit encoding} &  \multicolumn{2}{c}{qutrit encoding}\\ 
         type & 1 qubit & 2 qubit & 1 qutrit & 2 qutrit \\\hline \hline
         $\hat{U}^x$ & 15 & 3 & 5 & 0\\
         $(\hat{L}^z)^2$ & 1 & 0 & 2 & 0\\
         $\hat{L}^z\otimes\hat{L}^z$ & 4 & 8 & 4 & 3\\\hline
    \end{tabular}
    \caption{Gate Costs assuming native Qiskit gates compared to decompositions shown in Eqs. (\ref{eq:lz2rotdecomp}), (\ref{eq:uxrotdecomp}), and (\ref{eq:lzlzrotdecomp})}
    \label{tab:decomplz2}
\end{table}


\section{Simulation}
\label{sec:Simulation}
\subsection{State Preparation and Time Evolution}
\begin{figure}
    \centering
    \includegraphics[width=0.5\textwidth]{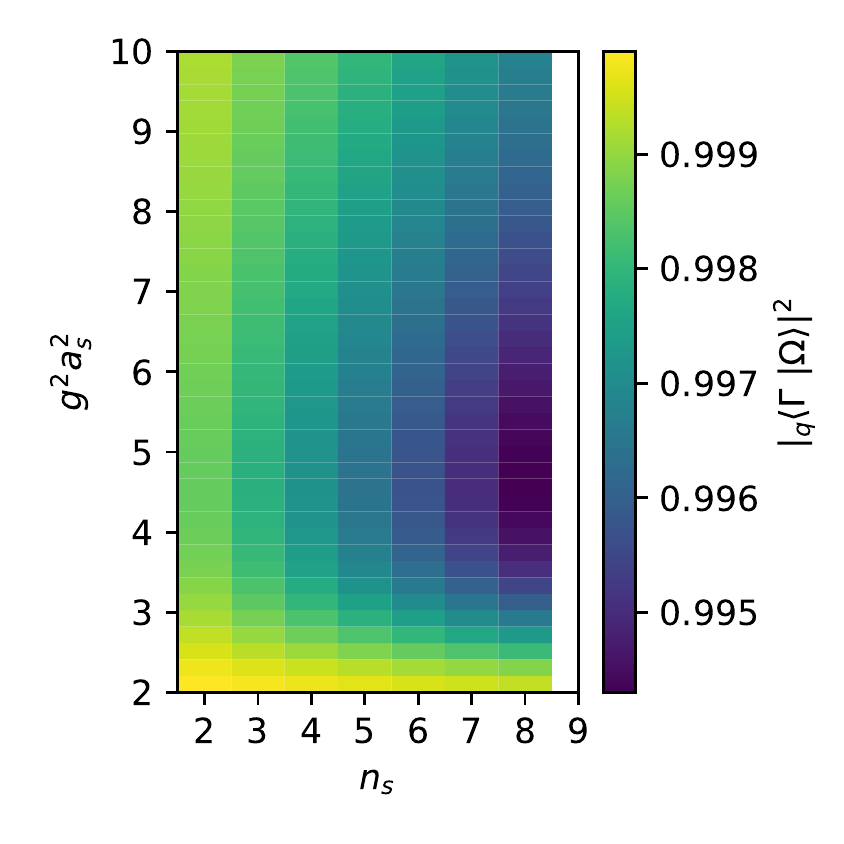}
    \caption{Overlap of the state $|1...1\rangle_q$ and exact ground state $|\Omega\rangle$ as a function of lattice size ($2 \leq n_s \leq 9$) and coupling strength $g^2 a_s^2$.}
    \label{fig:overlap}
\end{figure}

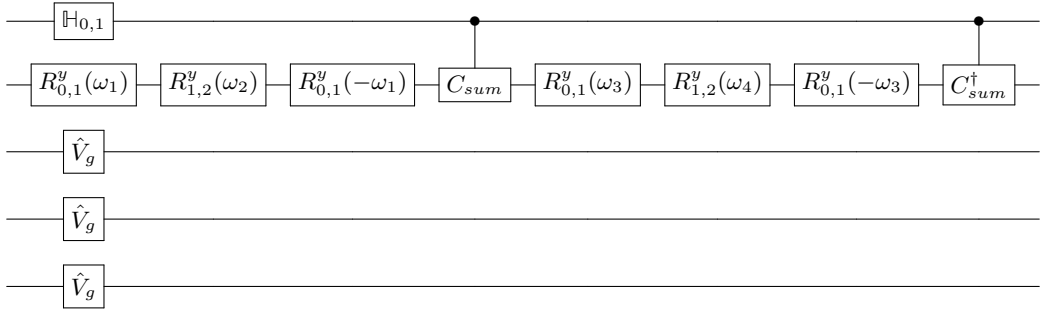
\begin{figure*}
\begin{gather*}
    \Qcircuit @C=1em @R=1em{
    & \gate{\mathbb{H}_{0, 1}} & \qw & \qw & \ctrl{1} & \qw & \qw & \qw & \ctrl{1} & \qw \\
    & \gate{R^y_{0, 1}(\omega_1)} & \gate{R^y_{1,2}(\omega_2)} & \gate{R^y_{0, 1}(-\omega_1)} & \gate{C_{sum}} & \gate{R^y_{0, 1}(\omega_3)} & \gate{R^y_{1,2}(\omega_4)} & \gate{R^y_{0, 1}(-\omega_3)} & \gate{C_{sum}^{\dagger}} & \qw \\
    & \gate{\hat{V}_g} & \qw & \qw & \qw  &  \qw & \qw & \qw & \qw & \qw \\
    & \gate{\hat{V}_g} & \qw & \qw & \qw  &  \qw & \qw & \qw & \qw & \qw \\
    & \gate{\hat{V}_g} & \qw & \qw & \qw  &  \qw & \qw & \qw & \qw & \qw \\
    }
\end{gather*}
\caption{Quantum circuit that creates the initial state, $|\Psi_0\rangle|\Psi_0\rangle|\Psi_0\rangle(|\Psi_0\rangle|0\rangle_a + 1/\mathcal{N}'\hat{U}^+|\Psi_0\rangle|1\rangle_a)$. $\mathbb{H}^{0, 1}$ is the Hadamard gate on the (0,1) subspace of the ancilla qutrit, $\hat{V}_g$ is defined in Eq. (\ref{eq:gsonsiteop}). The angles for $g^s a_s^2 = 5$ are $\omega_1 = -0.65273$, $\omega_2 = -1.43696$, $\omega_3 = 1.7837$ and $\omega_4 = 2.65568$.}
\label{fig:stateprep}
\end{figure*}

We will work in a regime $g^2 a_s ^2 = 5$ and $n_s = 4$. In this regime the ground state accurately represented by the iterative tensor product of the lowest eigenstate of the matrix
\begin{equation}
    \label{eq:onesiteop}
    \mathcal{A} = \frac{1}{2} \begin{pmatrix}
    g^2 a_s^2 + 1 & - 2 & 0\\
    -2 & 0 & -2\\
    0 & -2 & g^2 a_s^2 + 1\\
    \end{pmatrix},
\end{equation}
where we have set $Y = 1/2$ and $X=2$. The lowest energy eigenstate of this operator can be written
\begin{equation}
\label{eq:gsonsiteop}
\begin{split}
    |\Psi_0\rangle =& \frac{1}{\mathcal{N}} \Big(|0\rangle_q + b|1\rangle_q +  |2\rangle_q\Big)\\,
\end{split}
\end{equation}
where 
\begin{equation}
    \begin{split}
        \mathcal{N} &= \sqrt{2 + b^2}\\& \text{ and }\\
        b &= \frac{g^2 a_s^2 + 1 - \sqrt{(g^2 a_s^2 - 1)^2 + 32}}{4}.
    \end{split}
\end{equation}
The subscript $q$ indicates these are represented in the qutrit state values rather than the $\hat{L}^z$ spin values. The overlap of the state 
\begin{equation}
    \label{eq:groundstate}
    |\Gamma\rangle = \prod_{i=1}^{4} (|\Psi_0\rangle)^{\otimes}
\end{equation}
as a function of the couplings and lattice sizes is shown in Fig. \ref{fig:overlap}. 
We will demonstrate measuring the correlator,
\begin{equation}
    \label{eq:observable}
    \mathcal{C} = \langle \Gamma|e^{i t \hat{H}} \hat{U}^-_1 e^{-i t \hat{H}} \hat{U}^+_1|\Gamma\rangle.
\end{equation}
In order to measure this correlator, we need to be able to prepare the states $|\Gamma\rangle$ and $\hat{U}^+|\Gamma\rangle$. This is relatively straight forward to accomplish. The state $|\Psi_0\rangle$ in Eq. (\ref{eq:gsonsiteop}) can be prepared from the state $|0\rangle_q$ with two one qutrit rotations,
\begin{equation}
    \hat{V}_{g} = \hat{R}^y_{1, 2}(-\rho_2) \hat{R}^{y}_{0, 1}(\rho_1)
\end{equation}
where $\rho_1 = \text{arccos}(-1 / \mathcal{N})$ and $\rho_2 = \text{arcsin}(-1 / \sqrt{\mathcal{N}^2 - 1})$. Preparing the super position of $|\Psi_0\rangle$ and $\frac{1}{\mathcal{N}'}\hat{U}^+|\Psi_0\rangle$ is slightly harder. Using an ancilla, 2 $C_{sum}$ gates, and 7 one qutrit gates, we can prepare this superposition of states
\begin{equation}
    |\psi\rangle = \frac{1}{\sqrt{2}}\Bigg(|\Psi_0\rangle|0\rangle_a + \frac{1}{\mathcal{N}'}(|1\rangle + a |2\rangle)|1\rangle_a\Bigg).
\end{equation}
The circuit, $\hat{V}_{prep}$, which constructs this state is shown in Fig. \ref{fig:stateprep}.

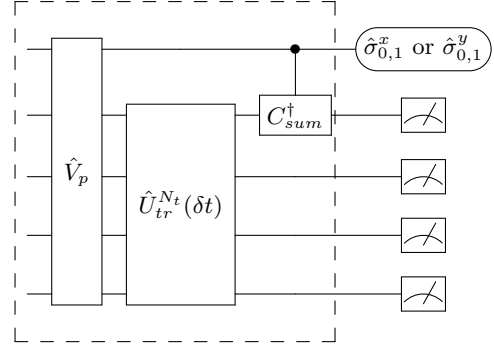
\begin{figure}
    \begin{gather*}
        \Qcircuit @C=1em @R=1em {
        & \multigate{4}{\hat{V}_p} & \qw & \ctrl{1} & \measure{\hat{\sigma}^x_{0, 1} \text{ or } \hat{\sigma}^y_{0, 1}}\\
        & \ghost{\hat{V}_p} & \multigate{3}{\hat{U}_{tr}^{N_t}(\delta t)} & \gate{C_{sum}^{\dagger}} & \meter \\
        & \ghost{\hat{V}_p} & \ghost{\hat{U}_{tr}^{N_t}(\delta t)} & \qw & \meter \\
        & \ghost{\hat{V}_p} & \ghost{\hat{U}_{tr}^{N_t}(\delta t)} & \qw & \meter \\
        & \ghost{\hat{V}_p} & \ghost{\hat{U}_{tr}^{N_t}(\delta t)} & \qw & \meter \gategroup{1}{2}{5}{4}{3em}{--}\\
        }
    \end{gather*}
    \caption{Circuit for measuring the correlator. The real part is found by measuring $\hat{\sigma}^x_{0,1}$ on the ancilla and the occupations on the on the first qubit. While the imaginary part is found by measuring $\hat{\sigma}^y_{0,1}$ on the ancilla instead. The boxed region will be defined as the operator $\hat{C}$.} 
    \label{fig:correlatorcircuit}
\end{figure}

After this it is relatively straight forward to measure the correlator. The Trotterized time evolution operator from Eq. (\ref{eq:trotter}) is applied to the working qubits followed by $C_{sum}^{\dagger}$ controlled on the ancilla to the first qutrit. Then by measuring the $\hat{\sigma}^x_{0, 1}$ on the ancilla and the states on the working qubits we extract the real part of the correlator. If $\hat{\sigma}^y_{0, 1}$ is measured on the ancilla, then the imaginary part of the correlator is extracted instead. Diagramatically this circuit is shown in Fig. \ref{fig:correlatorcircuit}.
The correlator is then given by the following quantum operations
\begin{equation}
    \label{eq:correlatorascircuits}
    \begin{split}
    \mathcal{C} &= (\langle |\hat{C}^{\dagger} (\hat{\sigma}^x_{0, 1})_a \hat{C}|\rangle + \langle |\hat{C}^{\dagger} (\hat{\sigma}^x_{0, 1})_a (\hat{Z}_2) \hat{C}|\rangle\\
    &+ i \langle |\hat{C}^{\dagger} (\hat{\sigma}^y_{0, 1})_a \hat{C}|\rangle + \langle |\hat{C}^{\dagger} (\hat{\sigma}^y_{0, 1})_a (\hat{Z}_2) \hat{C}|\rangle) / 2,
    \end{split}
\end{equation}
where $|\rangle$ represents the state $|0000\rangle\otimes(|0\rangle_a + |1\rangle_a)$.
The operator $\hat{Z}_2$ can be applied classically after measuring the qutrit state by apply a $-1$ to any measurement of the first working qutrit that is in the state $|2\rangle$. 

\subsection{Noisy Emulations}

\begin{figure*}
\centering
\includegraphics[width=0.95\textwidth]{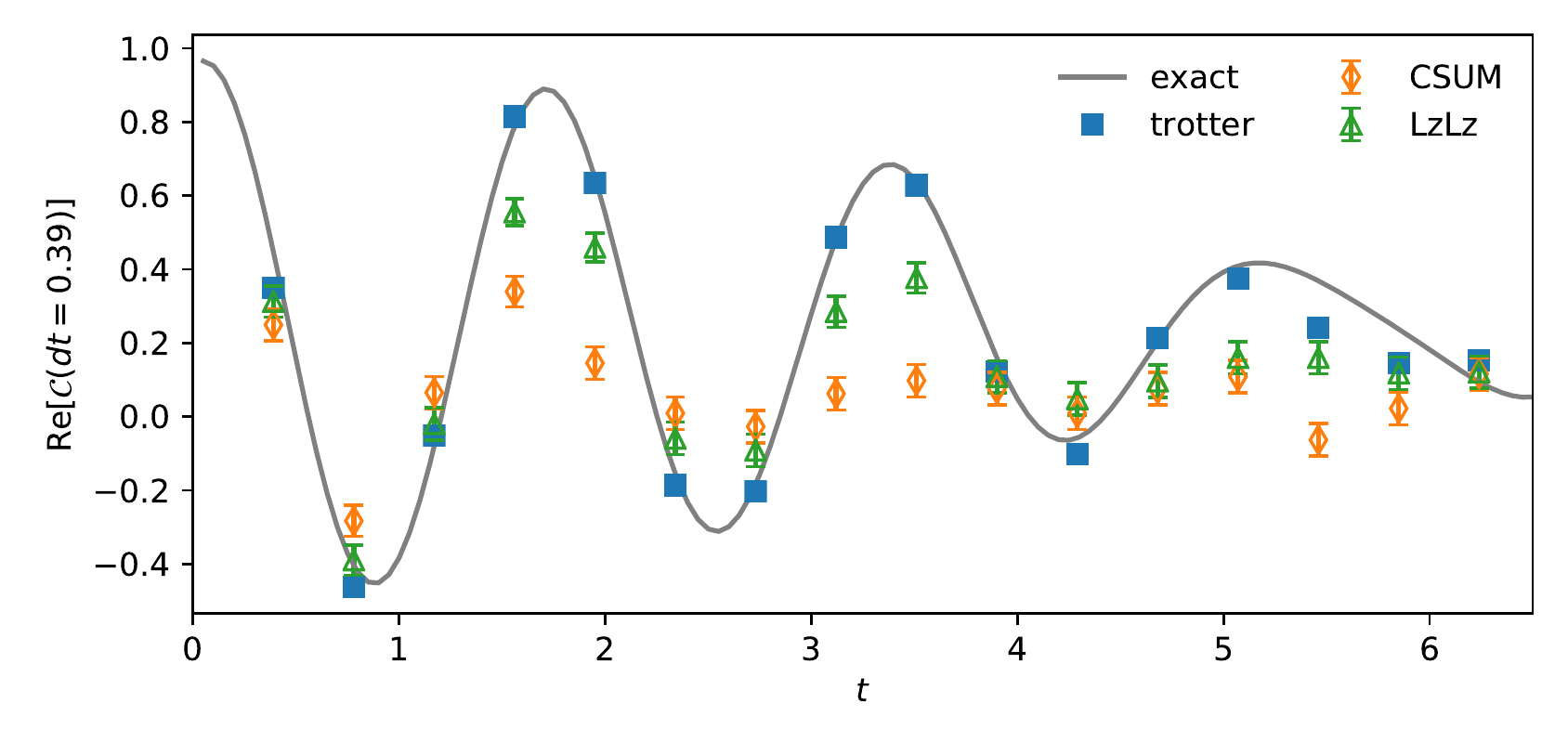}
\caption{Noisy Emulation of the real part of the Correlator $\mathcal{C}$ defined in Eq. (\ref{eq:observable}) for Trotter step size, $dt = 0.39$. Two different native gate sets were used: native $e^{-i \theta L^z\otimes L^z}$ plus one qutrit rotations from Eq. (\ref{eq:quditrotations}) and native $C_{sum}$ plus one qutrit rotations from Eq. (\ref{eq:quditrotations}). Only statistical errors are shown.}
\label{fig:noisy_simulation}
\end{figure*}

While many noise channels are available such as decay channels that correspond to spontaneous decays from higher excited states \cite{Miller_2018} and Pauli channels \cite{2006quant.ph.10127C} which correspond to ``bit" flips, phase flips. Here we use the following Pauli channel noise model for qutrit,
\begin{equation}
    \label{eq:paulinoise}
    \begin{split}
    \mathcal{E}(\rho; q) & = (1 - \sum_{i<j}\sum_{\alpha}(p^{\alpha}_{i, j})^q) \rho \\
    & + \sum_{i<j}\sum_{\alpha} (\sigma^{\alpha}_{i, j})^q \rho (\sigma^{\alpha}_{i, j})^q,
    \end{split}
\end{equation}
where $\alpha=x,y,z$, $(i, j)$ indicates the mixing between qutrit states $|i\rangle$ and $|j\rangle$, $p_{i, j}^{\alpha}$ is the probability of such error occurring, and $q$ indicates which qutrit to apply the noise operation on. This noise model is inspired by the way the fidelities of \cite{morvan2020qutrit} are reported. This should not be surprising as it looks like an extension of the qubit version of a Pauli noise model used by many \cite{GustafsonIsing,nielsen_chuang_2010,Kandala_2019,Kandala:2017aa}.  The two qutrit noise model is easily extendable from this using a tensor product of all the Pauli terms for the two qutrits.

Simulations using the noise model described in Eq. (\ref{eq:paulinoise}) are discussed here. The probabilities for the Pauli errors used are listed in Table \ref{tab:paulierrors} which were found for a recent randomized benchmark for a Transmon based qutrit system \cite{morvan2020qutrit}. The noise model was applied assuming that $R^x$, $R^y$ and two-qutrit rotations are noisy and that $R^z$ rotations are noiseless \cite{Blok:2020may}.

\begin{table}[t]
\centering
\begin{tabular}{cccc}
\hline\hline
Term & (0, 1) & (0, 2) & (1, 2) \\\hline
one qutrit term & 0.00038 & 0.00143 & 0.00068 \\
two qutrit term & 0.003 & 0.003 & 0.003 \\\hline\hline

\end{tabular}
\caption{Pauli errors for one and two qutrit Pauli terms. The two qutrit term applies to each element of the Pauli noise channel, i.e., each $\sigma^{\alpha}_{i, j} \sigma^{\beta}_{k, l}$ has that probability of occurring.}
\label{tab:paulierrors}
\end{table}

The results of noisy emulations of the observable $\mathcal{C}$ are shown in Fig. \ref{fig:noisy_simulation} for Trotter step size $\delta t = 0.39$ (additional simulations at $\delta t = 0.235$ and $0.31$ are shown in Appendix B). These steps sizes were chosen because they balanced the Trotter fidelity with the emulated noise in the gates to allow time dynamics to be observed. While the computer tested in \cite{morvan2020qutrit} had the controlled sum ($C_{sum}$) as the native two-qutrit gate, we also consider a case where a $e^{-i \theta L^z\otimes L^z}$ rotation can be used as a native gate assuming the same Pauli errors.

A clear feature is that the native $L^z L^z$ rotation allows for a 8 to 9 Trotter steps before the gate noise completely suppresses the signal while the native $C_{sum}$ allows for 4 to 5 Trotter steps before the signal is lost. This suggests that near term qutrit based machines such as those tested by \cite{morvan2020qutrit,Blok:2020may} may be able to simulate short term dynamics of this model and allow for early benchmarking of a more complicated field theory than the Transverse Ising Model. 

\section{Conclusions}
\label{sec:conclusions}
This work has shown how to implement the Abelian Higgs model on a qutrit based digital quantum computer. We show how to measure a two point correlation function and demonstrate that a few Trotter steps are feasible using current estimates for the Pauli channel noise on transmon qutrits \cite{morvan2020qutrit}. 

Simulations of this model on qutrit based quantum computers would be an significant step toward real time simulations of Quantum Electrodynamics and other field theories with continuous or larger symmetry groups. These simulations will help pave the way toward understanding the dynamics from an ab initio perspective inelastic scattering processes.

\begin{acknowledgments}
I would like to thank Henry Lamm, Judah Unmuth-Yockey, Jin Zhang, and Yannick Meurice for helpful comments on this work.
This work was supported by a Department of Energy Grant under Award Number DE-SC0019139.
\end{acknowledgments}

\appendix
\section{Rotation Decompositions}
\label{app:a}
\begin{table}[!ht]
    \centering
    \begin{tabular}{|c|c|c|c|}\hline
        angle & $a_{j, j+1}$ & $b_{j, j+1}$ & $c_{j, j+1}$ \\\hline
         $\alpha_{0, 1}$ & 2/3 & -1/3 & 0\\
         $\alpha_{1, 2}$ & 4/3 & -8/3 & 1\\
         $\alpha_{2, 3}$ & 2 & -7 & 5\\
         $\alpha_{3, 4}$ & 8/3 & -40 / 3 & 14\\
         $\alpha_{4, 5}$ & 10 / 3 & -65 / 3 & 30\\
         $\alpha_{5, 6}$ & 4 & -32 & 55 \\
         $\alpha_{6, 7}$ & 14/3 & 133 / 3& 91\\
         $\alpha_{7, 8}$ & 16/3 & -176/3 & 140\\
         $\alpha_{8, 9}$ & 6 & -75 & 204\\\hline
    \end{tabular}
    \caption{Angles $\alpha_{i, i +1}$ for the $e^{-i\theta(L^z)^2}$ rotations to provide rotations up to a spin truncation $n=9$. There is an anti-symmetry of the angles after passing the $\alpha_{(n - 1) / 2, (n - 1) / 2 + 1}$ where the angles are then follow the reverse pattern and are negative, e.g. for $n=1$ $\alpha_{0,1} = -\alpha_{1,2}$, and for $n=2$ $\alpha_{0, 1} = -\alpha_{3, 4}$ and $\alpha_{1, 2} = - \alpha_{2, 3}$.}
    \label{tab:coefficientslz2angles}
\end{table}
The coefficients $a_{j, j+1}$, $b_{j, j+1}$, and $c_{j, j+1}$ scale according to the following equations,
\begin{equation}
\label{eq:coefficientscaling}
\begin{split}
    a_{j, j+1} &= \frac{2}{3}j + \frac{2}{3}\\
    b_{j, j+1} &= -j^2 - \frac{4}{3} j - \frac{1}{3}\\
    c_{j, j+1} &= \frac{1}{3}j^3 +\frac{1}{2} j^2 +\frac{1}{6}j.\\
\end{split}
\end{equation}
Using Eq. (\ref{eq:coefficientscaling}), we can find an closed form expression for the angles $\alpha_{j, j+1}$,
\begin{equation}
    \label{eq:alphajj}
    \begin{split}
    \alpha_{j, j+1}(n) =& \frac{2j + 2}{3} n^2- \frac{3j^2 -4j -1}{3}n\\
    & + \frac{2j^3 + 3j^2 + j}{6}
    \end{split}
\end{equation}
\section{Additional Correlators}
\label{app:c}
\begin{figure*}
\centering
\includegraphics[width=0.95\textwidth]{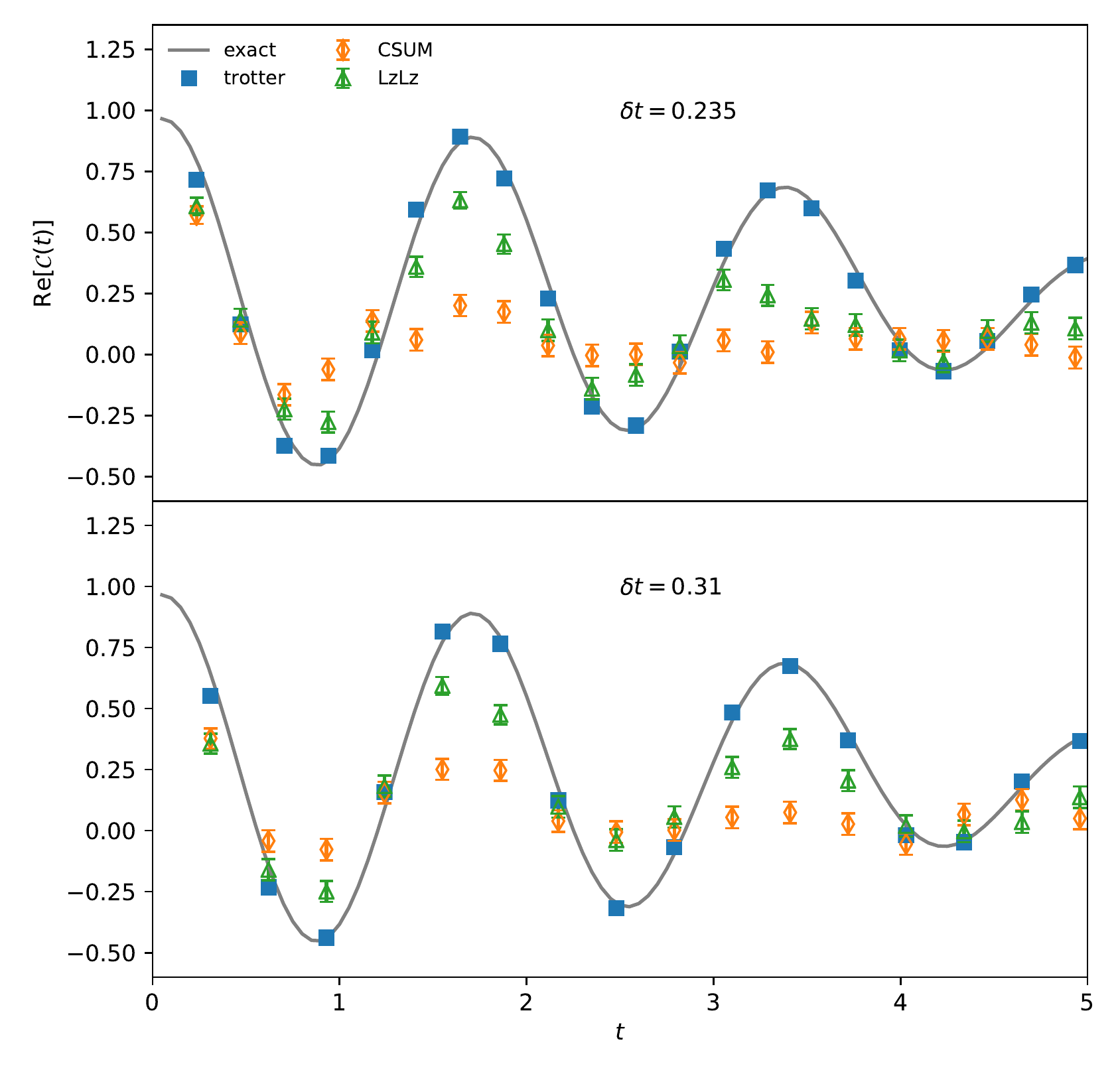}
\caption{Time evolution of the correlator $\mathcal{C}$ for 3 different Trotter step sizes, $\delta t = 0.235$ (left), $\delta t = 0.31$ (center), and $\delta t= 0.39$ (right). Using two different the native gate sets.}
\label{fig:noisy_simulationsmore}
\end{figure*}

Here we show the noisy simulations of the time evolution of $\mathcal{C}$ for 2 additional Trotter steps, $\delta t=0.235$ and $\delta t=0.31$. These Trotter step sizes were chosen because they are highly faithful as demonstrated in Fig. \ref{fig:noisy_simulationsmore}. In addition we can see a signal for these smaller Trotter step sizes using both native gate sets. 

\bibliography{main.bbl}

\begin{thebibliography}{89}%
\makeatletter
\providecommand \@ifxundefined [1]{%
 \@ifx{#1\undefined}
}%
\providecommand \@ifnum [1]{%
 \ifnum #1\expandafter \@firstoftwo
 \else \expandafter \@secondoftwo
 \fi
}%
\providecommand \@ifx [1]{%
 \ifx #1\expandafter \@firstoftwo
 \else \expandafter \@secondoftwo
 \fi
}%
\providecommand \natexlab [1]{#1}%
\providecommand \enquote  [1]{``#1''}%
\providecommand \bibnamefont  [1]{#1}%
\providecommand \bibfnamefont [1]{#1}%
\providecommand \citenamefont [1]{#1}%
\providecommand \href@noop [0]{\@secondoftwo}%
\providecommand \href [0]{\begingroup \@sanitize@url \@href}%
\providecommand \@href[1]{\@@startlink{#1}\@@href}%
\providecommand \@@href[1]{\endgroup#1\@@endlink}%
\providecommand \@sanitize@url [0]{\catcode `\\12\catcode `\$12\catcode
  `\&12\catcode `\#12\catcode `\^12\catcode `\_12\catcode `\%12\relax}%
\providecommand \@@startlink[1]{}%
\providecommand \@@endlink[0]{}%
\providecommand \url  [0]{\begingroup\@sanitize@url \@url }%
\providecommand \@url [1]{\endgroup\@href {#1}{\urlprefix }}%
\providecommand \urlprefix  [0]{URL }%
\providecommand \Eprint [0]{\href }%
\providecommand \doibase [0]{https://doi.org/}%
\providecommand \selectlanguage [0]{\@gobble}%
\providecommand \bibinfo  [0]{\@secondoftwo}%
\providecommand \bibfield  [0]{\@secondoftwo}%
\providecommand \translation [1]{[#1]}%
\providecommand \BibitemOpen [0]{}%
\providecommand \bibitemStop [0]{}%
\providecommand \bibitemNoStop [0]{.\EOS\space}%
\providecommand \EOS [0]{\spacefactor3000\relax}%
\providecommand \BibitemShut  [1]{\csname bibitem#1\endcsname}%
\let\auto@bib@innerbib\@empty
\bibitem [{dav(2020)}]{davoudi2020}%
  \BibitemOpen
  \bibfield  {title} {\bibinfo {title} {Recent progress in few-body physics},\
  }\bibfield  {journal} {\bibinfo  {journal} {Springer Proceedings in Physics}\
  }\href {https://doi.org/10.1007/978-3-030-32357-8}
  {10.1007/978-3-030-32357-8} (\bibinfo {year} {2020})\BibitemShut {NoStop}%
\bibitem [{\citenamefont {Padmanath}(2019)}]{padmanath2019hadron}%
  \BibitemOpen
  \bibfield  {author} {\bibinfo {author} {\bibfnamefont {M.}~\bibnamefont
  {Padmanath}},\ }\bibfield  {title} {\bibinfo {title} {Hadron spectroscopy and
  resonances: Review},\ }\href@noop {} {\  (\bibinfo {year} {2019})},\ \Eprint
  {https://arxiv.org/abs/1905.09651} {arXiv:1905.09651 [hep-lat]} \BibitemShut
  {NoStop}%
\bibitem [{\citenamefont {Alexandru}\ \emph {et~al.}(2016)\citenamefont
  {Alexandru}, \citenamefont {Basar}, \citenamefont {Bedaque}, \citenamefont
  {Vartak},\ and\ \citenamefont {Warrington}}]{Alexandru:2016gsd}%
  \BibitemOpen
  \bibfield  {author} {\bibinfo {author} {\bibfnamefont {A.}~\bibnamefont
  {Alexandru}}, \bibinfo {author} {\bibfnamefont {G.}~\bibnamefont {Basar}},
  \bibinfo {author} {\bibfnamefont {P.~F.}\ \bibnamefont {Bedaque}}, \bibinfo
  {author} {\bibfnamefont {S.}~\bibnamefont {Vartak}},\ and\ \bibinfo {author}
  {\bibfnamefont {N.~C.}\ \bibnamefont {Warrington}},\ }\bibfield  {title}
  {\bibinfo {title} {{Monte Carlo Study of Real Time Dynamics on the
  Lattice}},\ }\href {https://doi.org/10.1103/PhysRevLett.117.081602}
  {\bibfield  {journal} {\bibinfo  {journal} {Phys. Rev. Lett.}\ }\textbf
  {\bibinfo {volume} {117}},\ \bibinfo {pages} {081602} (\bibinfo {year}
  {2016})},\ \Eprint {https://arxiv.org/abs/1605.08040} {arXiv:1605.08040
  [hep-lat]} \BibitemShut {NoStop}%
\bibitem [{\citenamefont {Kanwar}\ and\ \citenamefont
  {Wagman}(2021)}]{Kanwar:2021tkd}%
  \BibitemOpen
  \bibfield  {author} {\bibinfo {author} {\bibfnamefont {G.}~\bibnamefont
  {Kanwar}}\ and\ \bibinfo {author} {\bibfnamefont {M.~L.}\ \bibnamefont
  {Wagman}},\ }\bibfield  {title} {\bibinfo {title} {{Real-time lattice gauge
  theory actions: unitarity, convergence, and path integral contour
  deformations}},\ }\href@noop {} {\  (\bibinfo {year} {2021})},\ \Eprint
  {https://arxiv.org/abs/2103.02602} {arXiv:2103.02602 [hep-lat]} \BibitemShut
  {NoStop}%
\bibitem [{\citenamefont {Jordan}\ \emph {et~al.}(2014)\citenamefont {Jordan},
  \citenamefont {Lee},\ and\ \citenamefont {Preskill}}]{Jordan:2011ci}%
  \BibitemOpen
  \bibfield  {author} {\bibinfo {author} {\bibfnamefont {S.~P.}\ \bibnamefont
  {Jordan}}, \bibinfo {author} {\bibfnamefont {K.~S.~M.}\ \bibnamefont {Lee}},\
  and\ \bibinfo {author} {\bibfnamefont {J.}~\bibnamefont {Preskill}},\
  }\bibfield  {title} {\bibinfo {title} {{Quantum Computation of Scattering in
  Scalar Quantum Field Theories}},\ }\href@noop {} {\bibfield  {journal}
  {\bibinfo  {journal} {Quant. Inf. Comput.}\ }\textbf {\bibinfo {volume}
  {14}},\ \bibinfo {pages} {1014} (\bibinfo {year} {2014})},\ \Eprint
  {https://arxiv.org/abs/1112.4833} {arXiv:1112.4833 [hep-th]} \BibitemShut
  {NoStop}%
\bibitem [{\citenamefont {Klco}\ and\ \citenamefont
  {Savage}(2019)}]{Klco_2019}%
  \BibitemOpen
  \bibfield  {author} {\bibinfo {author} {\bibfnamefont {N.}~\bibnamefont
  {Klco}}\ and\ \bibinfo {author} {\bibfnamefont {M.~J.}\ \bibnamefont
  {Savage}},\ }\bibfield  {title} {\bibinfo {title} {Digitization of scalar
  fields for quantum computing},\ }\bibfield  {journal} {\bibinfo  {journal}
  {Physical Review A}\ }\textbf {\bibinfo {volume} {99}},\ \href
  {https://doi.org/10.1103/physreva.99.052335} {10.1103/physreva.99.052335}
  (\bibinfo {year} {2019})\BibitemShut {NoStop}%
\bibitem [{\citenamefont {Bazavov}\ \emph
  {et~al.}(2015{\natexlab{a}})\citenamefont {Bazavov}, \citenamefont {Meurice},
  \citenamefont {Tsai}, \citenamefont {Unmuth-Yockey},\ and\ \citenamefont
  {Zhang}}]{Bazavov_2015}%
  \BibitemOpen
  \bibfield  {author} {\bibinfo {author} {\bibfnamefont {A.}~\bibnamefont
  {Bazavov}}, \bibinfo {author} {\bibfnamefont {Y.}~\bibnamefont {Meurice}},
  \bibinfo {author} {\bibfnamefont {S.-W.}\ \bibnamefont {Tsai}}, \bibinfo
  {author} {\bibfnamefont {J.}~\bibnamefont {Unmuth-Yockey}},\ and\ \bibinfo
  {author} {\bibfnamefont {J.}~\bibnamefont {Zhang}},\ }\bibfield  {title}
  {\bibinfo {title} {Gauge-invariant implementation of the abelian-higgs model
  on optical lattices},\ }\bibfield  {journal} {\bibinfo  {journal} {Physical
  Review D}\ }\textbf {\bibinfo {volume} {92}},\ \href
  {https://doi.org/10.1103/physrevd.92.076003} {10.1103/physrevd.92.076003}
  (\bibinfo {year} {2015}{\natexlab{a}})\BibitemShut {NoStop}%
\bibitem [{\citenamefont {Bazavov}\ \emph
  {et~al.}(2015{\natexlab{b}})\citenamefont {Bazavov}, \citenamefont {Meurice},
  \citenamefont {Tsai}, \citenamefont {Unmuth-Yockey},\ and\ \citenamefont
  {Zhang}}]{bazavov2015effective}%
  \BibitemOpen
  \bibfield  {author} {\bibinfo {author} {\bibfnamefont {A.}~\bibnamefont
  {Bazavov}}, \bibinfo {author} {\bibfnamefont {Y.}~\bibnamefont {Meurice}},
  \bibinfo {author} {\bibfnamefont {S.-W.}\ \bibnamefont {Tsai}}, \bibinfo
  {author} {\bibfnamefont {J.}~\bibnamefont {Unmuth-Yockey}},\ and\ \bibinfo
  {author} {\bibfnamefont {J.}~\bibnamefont {Zhang}},\ }\href@noop {} {\bibinfo
  {title} {Effective action for the abelian-higgs model for a gauge-invariant
  implementation on optical lattices}} (\bibinfo {year} {2015}{\natexlab{b}}),\
  \Eprint {https://arxiv.org/abs/1512.01737} {arXiv:1512.01737 [hep-lat]}
  \BibitemShut {NoStop}%
\bibitem [{\citenamefont {Zhang}\ \emph {et~al.}(2018)\citenamefont {Zhang},
  \citenamefont {Unmuth-Yockey}, \citenamefont {Zeiher}, \citenamefont
  {Bazavov}, \citenamefont {Tsai},\ and\ \citenamefont
  {Meurice}}]{PhysRevLett.121.223201}%
  \BibitemOpen
  \bibfield  {author} {\bibinfo {author} {\bibfnamefont {J.}~\bibnamefont
  {Zhang}}, \bibinfo {author} {\bibfnamefont {J.}~\bibnamefont
  {Unmuth-Yockey}}, \bibinfo {author} {\bibfnamefont {J.}~\bibnamefont
  {Zeiher}}, \bibinfo {author} {\bibfnamefont {A.}~\bibnamefont {Bazavov}},
  \bibinfo {author} {\bibfnamefont {S.-W.}\ \bibnamefont {Tsai}},\ and\
  \bibinfo {author} {\bibfnamefont {Y.}~\bibnamefont {Meurice}},\ }\bibfield
  {title} {\bibinfo {title} {Quantum simulation of the universal features of
  the polyakov loop},\ }\href {https://doi.org/10.1103/PhysRevLett.121.223201}
  {\bibfield  {journal} {\bibinfo  {journal} {Phys. Rev. Lett.}\ }\textbf
  {\bibinfo {volume} {121}},\ \bibinfo {pages} {223201} (\bibinfo {year}
  {2018})}\BibitemShut {NoStop}%
\bibitem [{\citenamefont {Unmuth-Yockey}\ \emph {et~al.}(2018)\citenamefont
  {Unmuth-Yockey}, \citenamefont {Zhang}, \citenamefont {Bazavov},
  \citenamefont {Meurice},\ and\ \citenamefont {Tsai}}]{Unmuth_Yockey_2018}%
  \BibitemOpen
  \bibfield  {author} {\bibinfo {author} {\bibfnamefont {J.}~\bibnamefont
  {Unmuth-Yockey}}, \bibinfo {author} {\bibfnamefont {J.}~\bibnamefont
  {Zhang}}, \bibinfo {author} {\bibfnamefont {A.}~\bibnamefont {Bazavov}},
  \bibinfo {author} {\bibfnamefont {Y.}~\bibnamefont {Meurice}},\ and\ \bibinfo
  {author} {\bibfnamefont {S.-W.}\ \bibnamefont {Tsai}},\ }\bibfield  {title}
  {\bibinfo {title} {Universal features of the abelian polyakov loop in 1+1
  dimensions},\ }\bibfield  {journal} {\bibinfo  {journal} {Physical Review D}\
  }\textbf {\bibinfo {volume} {98}},\ \href
  {https://doi.org/10.1103/physrevd.98.094511} {10.1103/physrevd.98.094511}
  (\bibinfo {year} {2018})\BibitemShut {NoStop}%
\bibitem [{Mus(2017)}]{Muschik_2017}%
  \BibitemOpen
  \bibfield  {title} {\bibinfo {title} {U(1) wilson lattice gauge theories in
  digital quantum simulators},\ }\href
  {https://doi.org/10.1088/1367-2630/aa89ab} {\bibfield  {journal} {\bibinfo
  {journal} {New Journal of Physics}\ }\textbf {\bibinfo {volume} {19}},\
  \bibinfo {pages} {103020} (\bibinfo {year} {2017})}\BibitemShut {NoStop}%
\bibitem [{\citenamefont {Shaw}\ \emph {et~al.}(2020)\citenamefont {Shaw},
  \citenamefont {Lougovski}, \citenamefont {Stryker},\ and\ \citenamefont
  {Wiebe}}]{Shaw_2020}%
  \BibitemOpen
  \bibfield  {author} {\bibinfo {author} {\bibfnamefont {A.~F.}\ \bibnamefont
  {Shaw}}, \bibinfo {author} {\bibfnamefont {P.}~\bibnamefont {Lougovski}},
  \bibinfo {author} {\bibfnamefont {J.~R.}\ \bibnamefont {Stryker}},\ and\
  \bibinfo {author} {\bibfnamefont {N.}~\bibnamefont {Wiebe}},\ }\bibfield
  {title} {\bibinfo {title} {Quantum algorithms for simulating the lattice
  schwinger model},\ }\href {https://doi.org/10.22331/q-2020-08-10-306}
  {\bibfield  {journal} {\bibinfo  {journal} {Quantum}\ }\textbf {\bibinfo
  {volume} {4}},\ \bibinfo {pages} {306} (\bibinfo {year} {2020})}\BibitemShut
  {NoStop}%
\bibitem [{\citenamefont {Kaplan}\ and\ \citenamefont
  {Stryker}(2020)}]{Kaplan_2020}%
  \BibitemOpen
  \bibfield  {author} {\bibinfo {author} {\bibfnamefont {D.~B.}\ \bibnamefont
  {Kaplan}}\ and\ \bibinfo {author} {\bibfnamefont {J.~R.}\ \bibnamefont
  {Stryker}},\ }\bibfield  {title} {\bibinfo {title} {Gauss’s law, duality,
  and the hamiltonian formulation of u(1) lattice gauge theory},\ }\bibfield
  {journal} {\bibinfo  {journal} {Physical Review D}\ }\textbf {\bibinfo
  {volume} {102}},\ \href {https://doi.org/10.1103/physrevd.102.094515}
  {10.1103/physrevd.102.094515} (\bibinfo {year} {2020})\BibitemShut {NoStop}%
\bibitem [{\citenamefont {Raychowdhury}\ and\ \citenamefont
  {Stryker}(2020)}]{Raychowdhury_2020}%
  \BibitemOpen
  \bibfield  {author} {\bibinfo {author} {\bibfnamefont {I.}~\bibnamefont
  {Raychowdhury}}\ and\ \bibinfo {author} {\bibfnamefont {J.~R.}\ \bibnamefont
  {Stryker}},\ }\bibfield  {title} {\bibinfo {title} {Loop, string, and hadron
  dynamics in su(2) hamiltonian lattice gauge theories},\ }\bibfield  {journal}
  {\bibinfo  {journal} {Physical Review D}\ }\textbf {\bibinfo {volume}
  {101}},\ \href {https://doi.org/10.1103/physrevd.101.114502}
  {10.1103/physrevd.101.114502} (\bibinfo {year} {2020})\BibitemShut {NoStop}%
\bibitem [{\citenamefont {Alexandru}\ \emph {et~al.}(2019)\citenamefont
  {Alexandru}, \citenamefont {Bedaque}, \citenamefont {Harmalkar},
  \citenamefont {Lamm}, \citenamefont {Lawrence},\ and\ \citenamefont
  {Warrington}}]{Alexandru_2019}%
  \BibitemOpen
  \bibfield  {author} {\bibinfo {author} {\bibfnamefont {A.}~\bibnamefont
  {Alexandru}}, \bibinfo {author} {\bibfnamefont {P.~F.}\ \bibnamefont
  {Bedaque}}, \bibinfo {author} {\bibfnamefont {S.}~\bibnamefont {Harmalkar}},
  \bibinfo {author} {\bibfnamefont {H.}~\bibnamefont {Lamm}}, \bibinfo {author}
  {\bibfnamefont {S.}~\bibnamefont {Lawrence}},\ and\ \bibinfo {author}
  {\bibfnamefont {N.~C.}\ \bibnamefont {Warrington}},\ }\bibfield  {title}
  {\bibinfo {title} {Gluon field digitization for quantum computers},\
  }\bibfield  {journal} {\bibinfo  {journal} {Physical Review D}\ }\textbf
  {\bibinfo {volume} {100}},\ \href
  {https://doi.org/10.1103/physrevd.100.114501} {10.1103/physrevd.100.114501}
  (\bibinfo {year} {2019})\BibitemShut {NoStop}%
\bibitem [{\citenamefont {Hackett}\ \emph {et~al.}(2019)\citenamefont
  {Hackett}, \citenamefont {Howe}, \citenamefont {Hughes}, \citenamefont {Jay},
  \citenamefont {Neil},\ and\ \citenamefont {Simone}}]{Hackett_2019}%
  \BibitemOpen
  \bibfield  {author} {\bibinfo {author} {\bibfnamefont {D.~C.}\ \bibnamefont
  {Hackett}}, \bibinfo {author} {\bibfnamefont {K.}~\bibnamefont {Howe}},
  \bibinfo {author} {\bibfnamefont {C.}~\bibnamefont {Hughes}}, \bibinfo
  {author} {\bibfnamefont {W.}~\bibnamefont {Jay}}, \bibinfo {author}
  {\bibfnamefont {E.~T.}\ \bibnamefont {Neil}},\ and\ \bibinfo {author}
  {\bibfnamefont {J.~N.}\ \bibnamefont {Simone}},\ }\bibfield  {title}
  {\bibinfo {title} {Digitizing gauge fields: Lattice monte carlo results for
  future quantum computers},\ }\bibfield  {journal} {\bibinfo  {journal}
  {Physical Review A}\ }\textbf {\bibinfo {volume} {99}},\ \href
  {https://doi.org/10.1103/physreva.99.062341} {10.1103/physreva.99.062341}
  (\bibinfo {year} {2019})\BibitemShut {NoStop}%
\bibitem [{\citenamefont {Ji}\ \emph {et~al.}(2020)\citenamefont {Ji},
  \citenamefont {Lamm},\ and\ \citenamefont {Zhu}}]{Ji_2020}%
  \BibitemOpen
  \bibfield  {author} {\bibinfo {author} {\bibfnamefont {Y.}~\bibnamefont
  {Ji}}, \bibinfo {author} {\bibfnamefont {H.}~\bibnamefont {Lamm}},\ and\
  \bibinfo {author} {\bibfnamefont {S.}~\bibnamefont {Zhu}},\ }\bibfield
  {title} {\bibinfo {title} {Gluon field digitization via group space
  decimation for quantum computers},\ }\bibfield  {journal} {\bibinfo
  {journal} {Physical Review D}\ }\textbf {\bibinfo {volume} {102}},\ \href
  {https://doi.org/10.1103/physrevd.102.114513} {10.1103/physrevd.102.114513}
  (\bibinfo {year} {2020})\BibitemShut {NoStop}%
\bibitem [{\citenamefont {Ciavarella}\ \emph {et~al.}(2021)\citenamefont
  {Ciavarella}, \citenamefont {Klco},\ and\ \citenamefont
  {Savage}}]{ciavarella2021trailhead}%
  \BibitemOpen
  \bibfield  {author} {\bibinfo {author} {\bibfnamefont {A.}~\bibnamefont
  {Ciavarella}}, \bibinfo {author} {\bibfnamefont {N.}~\bibnamefont {Klco}},\
  and\ \bibinfo {author} {\bibfnamefont {M.~J.}\ \bibnamefont {Savage}},\
  }\href@noop {} {\bibinfo {title} {A trailhead for quantum simulation of su(3)
  yang-mills lattice gauge theory in the local multiplet basis}} (\bibinfo
  {year} {2021}),\ \Eprint {https://arxiv.org/abs/2101.10227} {arXiv:2101.10227
  [quant-ph]} \BibitemShut {NoStop}%
\bibitem [{\citenamefont {Klco}\ \emph {et~al.}(2020)\citenamefont {Klco},
  \citenamefont {Savage},\ and\ \citenamefont {Stryker}}]{klco_2020}%
  \BibitemOpen
  \bibfield  {author} {\bibinfo {author} {\bibfnamefont {N.}~\bibnamefont
  {Klco}}, \bibinfo {author} {\bibfnamefont {M.~J.}\ \bibnamefont {Savage}},\
  and\ \bibinfo {author} {\bibfnamefont {J.~R.}\ \bibnamefont {Stryker}},\
  }\bibfield  {title} {\bibinfo {title} {Su(2) non-abelian gauge field theory
  in one dimension on digital quantum computers},\ }\bibfield  {journal}
  {\bibinfo  {journal} {Physical Review D}\ }\textbf {\bibinfo {volume}
  {101}},\ \href {https://doi.org/10.1103/physrevd.101.074512}
  {10.1103/physrevd.101.074512} (\bibinfo {year} {2020})\BibitemShut {NoStop}%
\bibitem [{\citenamefont {Chandrasekharan}\ and\ \citenamefont
  {Wiese}(1997)}]{Chandrasekharan:1996ih}%
  \BibitemOpen
  \bibfield  {author} {\bibinfo {author} {\bibfnamefont {S.}~\bibnamefont
  {Chandrasekharan}}\ and\ \bibinfo {author} {\bibfnamefont {U.~J.}\
  \bibnamefont {Wiese}},\ }\bibfield  {title} {\bibinfo {title} {{Quantum link
  models: A Discrete approach to gauge theories}},\ }\href
  {https://doi.org/10.1016/S0550-3213(97)00006-0} {\bibfield  {journal}
  {\bibinfo  {journal} {Nucl. Phys. B}\ }\textbf {\bibinfo {volume} {492}},\
  \bibinfo {pages} {455} (\bibinfo {year} {1997})},\ \Eprint
  {https://arxiv.org/abs/hep-lat/9609042} {arXiv:hep-lat/9609042} \BibitemShut
  {NoStop}%
\bibitem [{\citenamefont {Brower}\ \emph {et~al.}(1999)\citenamefont {Brower},
  \citenamefont {Chandrasekharan},\ and\ \citenamefont
  {Wiese}}]{Brower:1997ha}%
  \BibitemOpen
  \bibfield  {author} {\bibinfo {author} {\bibfnamefont {R.}~\bibnamefont
  {Brower}}, \bibinfo {author} {\bibfnamefont {S.}~\bibnamefont
  {Chandrasekharan}},\ and\ \bibinfo {author} {\bibfnamefont {U.~J.}\
  \bibnamefont {Wiese}},\ }\bibfield  {title} {\bibinfo {title} {{QCD as a
  quantum link model}},\ }\href {https://doi.org/10.1103/PhysRevD.60.094502}
  {\bibfield  {journal} {\bibinfo  {journal} {Phys. Rev.}\ }\textbf {\bibinfo
  {volume} {D60}},\ \bibinfo {pages} {094502} (\bibinfo {year} {1999})},\
  \Eprint {https://arxiv.org/abs/hep-th/9704106} {arXiv:hep-th/9704106
  [hep-th]} \BibitemShut {NoStop}%
\bibitem [{\citenamefont {Beard}\ \emph {et~al.}(1998)\citenamefont {Beard},
  \citenamefont {Brower}, \citenamefont {Chandrasekharan}, \citenamefont
  {Chen}, \citenamefont {Tsapalis},\ and\ \citenamefont {Wiese}}]{Beard_1998}%
  \BibitemOpen
  \bibfield  {author} {\bibinfo {author} {\bibfnamefont {B.}~\bibnamefont
  {Beard}}, \bibinfo {author} {\bibfnamefont {R.}~\bibnamefont {Brower}},
  \bibinfo {author} {\bibfnamefont {S.}~\bibnamefont {Chandrasekharan}},
  \bibinfo {author} {\bibfnamefont {D.}~\bibnamefont {Chen}}, \bibinfo {author}
  {\bibfnamefont {A.}~\bibnamefont {Tsapalis}},\ and\ \bibinfo {author}
  {\bibfnamefont {U.-J.}\ \bibnamefont {Wiese}},\ }\bibfield  {title} {\bibinfo
  {title} {D-theory: field theory via dimensional reduction of discrete
  variables},\ }\href {https://doi.org/10.1016/s0920-5632(97)00900-6}
  {\bibfield  {journal} {\bibinfo  {journal} {Nuclear Physics B - Proceedings
  Supplements}\ }\textbf {\bibinfo {volume} {63}},\ \bibinfo {pages}
  {775–789} (\bibinfo {year} {1998})}\BibitemShut {NoStop}%
\bibitem [{\citenamefont {Brower}\ \emph {et~al.}(2004)\citenamefont {Brower},
  \citenamefont {Chandrasekharan}, \citenamefont {Riederer},\ and\
  \citenamefont {Wiese}}]{BROWER2004149}%
  \BibitemOpen
  \bibfield  {author} {\bibinfo {author} {\bibfnamefont {R.}~\bibnamefont
  {Brower}}, \bibinfo {author} {\bibfnamefont {S.}~\bibnamefont
  {Chandrasekharan}}, \bibinfo {author} {\bibfnamefont {S.}~\bibnamefont
  {Riederer}},\ and\ \bibinfo {author} {\bibfnamefont {U.-J.}\ \bibnamefont
  {Wiese}},\ }\bibfield  {title} {\bibinfo {title} {D-theory: field
  quantization by dimensional reduction of discrete variables},\ }\href
  {https://doi.org/https://doi.org/10.1016/j.nuclphysb.2004.06.007} {\bibfield
  {journal} {\bibinfo  {journal} {Nuclear Physics B}\ }\textbf {\bibinfo
  {volume} {693}},\ \bibinfo {pages} {149} (\bibinfo {year}
  {2004})}\BibitemShut {NoStop}%
\bibitem [{\citenamefont {Zohar}\ \emph {et~al.}(2013)\citenamefont {Zohar},
  \citenamefont {Cirac},\ and\ \citenamefont {Reznik}}]{Zohar_2013}%
  \BibitemOpen
  \bibfield  {author} {\bibinfo {author} {\bibfnamefont {E.}~\bibnamefont
  {Zohar}}, \bibinfo {author} {\bibfnamefont {J.~I.}\ \bibnamefont {Cirac}},\
  and\ \bibinfo {author} {\bibfnamefont {B.}~\bibnamefont {Reznik}},\
  }\bibfield  {title} {\bibinfo {title} {Quantum simulations of gauge theories
  with ultracold atoms: Local gauge invariance from angular-momentum
  conservation},\ }\bibfield  {journal} {\bibinfo  {journal} {Physical Review
  A}\ }\textbf {\bibinfo {volume} {88}},\ \href
  {https://doi.org/10.1103/physreva.88.023617} {10.1103/physreva.88.023617}
  (\bibinfo {year} {2013})\BibitemShut {NoStop}%
\bibitem [{\citenamefont {{Haase}}\ \emph {et~al.}(2020)\citenamefont
  {{Haase}}, \citenamefont {{Dellantonio}}, \citenamefont {{Celi}},
  \citenamefont {{Paulson}}, \citenamefont {{Kan}}, \citenamefont {{Jansen}},\
  and\ \citenamefont {{Muschik}}}]{2020arXiv200614160H}%
  \BibitemOpen
  \bibfield  {author} {\bibinfo {author} {\bibfnamefont {J.~F.}\ \bibnamefont
  {{Haase}}}, \bibinfo {author} {\bibfnamefont {L.}~\bibnamefont
  {{Dellantonio}}}, \bibinfo {author} {\bibfnamefont {A.}~\bibnamefont
  {{Celi}}}, \bibinfo {author} {\bibfnamefont {D.}~\bibnamefont {{Paulson}}},
  \bibinfo {author} {\bibfnamefont {A.}~\bibnamefont {{Kan}}}, \bibinfo
  {author} {\bibfnamefont {K.}~\bibnamefont {{Jansen}}},\ and\ \bibinfo
  {author} {\bibfnamefont {C.~A.}\ \bibnamefont {{Muschik}}},\ }\bibfield
  {title} {\bibinfo {title} {{A resource efficient approach for quantum and
  classical simulations of gauge theories in particle physics}},\ }\href@noop
  {} {\bibfield  {journal} {\bibinfo  {journal} {arXiv e-prints}\ ,\ \bibinfo
  {eid} {arXiv:2006.14160}} (\bibinfo {year} {2020})},\ \Eprint
  {https://arxiv.org/abs/2006.14160} {arXiv:2006.14160 [quant-ph]} \BibitemShut
  {NoStop}%
\bibitem [{\citenamefont {Hasenfratz}\ and\ \citenamefont
  {Niedermayer}(2001)}]{Hasenfratz:2001iz}%
  \BibitemOpen
  \bibfield  {author} {\bibinfo {author} {\bibfnamefont {P.}~\bibnamefont
  {Hasenfratz}}\ and\ \bibinfo {author} {\bibfnamefont {F.}~\bibnamefont
  {Niedermayer}},\ }\bibfield  {title} {\bibinfo {title} {{Asymptotic freedom
  with discrete spin variables?}},\ }\href
  {https://doi.org/10.22323/1.007.0229} {\bibfield  {journal} {\bibinfo
  {journal} {PoS}\ }\textbf {\bibinfo {volume} {HEP2001}},\ \bibinfo {pages}
  {229} (\bibinfo {year} {2001})},\ \Eprint
  {https://arxiv.org/abs/hep-lat/0112003} {arXiv:hep-lat/0112003} \BibitemShut
  {NoStop}%
\bibitem [{\citenamefont {Caracciolo}\ \emph
  {et~al.}(2001{\natexlab{a}})\citenamefont {Caracciolo}, \citenamefont
  {Montanari},\ and\ \citenamefont {Pelissetto}}]{Caracciolo_2001}%
  \BibitemOpen
  \bibfield  {author} {\bibinfo {author} {\bibfnamefont {S.}~\bibnamefont
  {Caracciolo}}, \bibinfo {author} {\bibfnamefont {A.}~\bibnamefont
  {Montanari}},\ and\ \bibinfo {author} {\bibfnamefont {A.}~\bibnamefont
  {Pelissetto}},\ }\bibfield  {title} {\bibinfo {title} {Asymptotically free
  models and discrete non-abelian groups},\ }\href
  {https://doi.org/10.1016/s0370-2693(01)00674-8} {\bibfield  {journal}
  {\bibinfo  {journal} {Physics Letters B}\ }\textbf {\bibinfo {volume}
  {513}},\ \bibinfo {pages} {223–231} (\bibinfo {year}
  {2001}{\natexlab{a}})}\BibitemShut {NoStop}%
\bibitem [{\citenamefont {Patrascioiu}\ and\ \citenamefont
  {Seiler}(1998)}]{PhysRevE.57.111}%
  \BibitemOpen
  \bibfield  {author} {\bibinfo {author} {\bibfnamefont {A.}~\bibnamefont
  {Patrascioiu}}\ and\ \bibinfo {author} {\bibfnamefont {E.}~\bibnamefont
  {Seiler}},\ }\bibfield  {title} {\bibinfo {title} {Continuum limit of
  two-dimensional spin models with continuous symmetry and conformal quantum
  field theory},\ }\href {https://doi.org/10.1103/PhysRevE.57.111} {\bibfield
  {journal} {\bibinfo  {journal} {Phys. Rev. E}\ }\textbf {\bibinfo {volume}
  {57}},\ \bibinfo {pages} {111} (\bibinfo {year} {1998})}\BibitemShut
  {NoStop}%
\bibitem [{\citenamefont {Krcmar}\ \emph {et~al.}(2016)\citenamefont {Krcmar},
  \citenamefont {Gendiar},\ and\ \citenamefont {Nishino}}]{PhysRevE.94.022134}%
  \BibitemOpen
  \bibfield  {author} {\bibinfo {author} {\bibfnamefont {R.}~\bibnamefont
  {Krcmar}}, \bibinfo {author} {\bibfnamefont {A.}~\bibnamefont {Gendiar}},\
  and\ \bibinfo {author} {\bibfnamefont {T.}~\bibnamefont {Nishino}},\
  }\bibfield  {title} {\bibinfo {title} {Phase diagram of a truncated
  tetrahedral model},\ }\href {https://doi.org/10.1103/PhysRevE.94.022134}
  {\bibfield  {journal} {\bibinfo  {journal} {Phys. Rev. E}\ }\textbf {\bibinfo
  {volume} {94}},\ \bibinfo {pages} {022134} (\bibinfo {year}
  {2016})}\BibitemShut {NoStop}%
\bibitem [{\citenamefont {Caracciolo}\ \emph
  {et~al.}(2001{\natexlab{b}})\citenamefont {Caracciolo}, \citenamefont
  {Montanari},\ and\ \citenamefont {Pelissetto}}]{CARACCIOLO2001223}%
  \BibitemOpen
  \bibfield  {author} {\bibinfo {author} {\bibfnamefont {S.}~\bibnamefont
  {Caracciolo}}, \bibinfo {author} {\bibfnamefont {A.}~\bibnamefont
  {Montanari}},\ and\ \bibinfo {author} {\bibfnamefont {A.}~\bibnamefont
  {Pelissetto}},\ }\bibfield  {title} {\bibinfo {title} {Asymptotically free
  models and discrete non-abelian groups},\ }\href
  {https://doi.org/https://doi.org/10.1016/S0370-2693(01)00674-8} {\bibfield
  {journal} {\bibinfo  {journal} {Physics Letters B}\ }\textbf {\bibinfo
  {volume} {513}},\ \bibinfo {pages} {223} (\bibinfo {year}
  {2001}{\natexlab{b}})}\BibitemShut {NoStop}%
\bibitem [{\citenamefont {Unmuth-Yockey}(2019)}]{Unmuth_Yockey_2019}%
  \BibitemOpen
  \bibfield  {author} {\bibinfo {author} {\bibfnamefont {J.~F.}\ \bibnamefont
  {Unmuth-Yockey}},\ }\bibfield  {title} {\bibinfo {title} {Gauge-invariant
  rotor hamiltonian from dual variables of 3d u(1) gauge theory},\ }\bibfield
  {journal} {\bibinfo  {journal} {Physical Review D}\ }\textbf {\bibinfo
  {volume} {99}},\ \href {https://doi.org/10.1103/physrevd.99.074502}
  {10.1103/physrevd.99.074502} (\bibinfo {year} {2019})\BibitemShut {NoStop}%
\bibitem [{\citenamefont {Clark}\ \emph {et~al.}(2009)\citenamefont {Clark},
  \citenamefont {Metodi}, \citenamefont {Gasster},\ and\ \citenamefont
  {Brown}}]{PhysRevA.79.062314}%
  \BibitemOpen
  \bibfield  {author} {\bibinfo {author} {\bibfnamefont {C.~R.}\ \bibnamefont
  {Clark}}, \bibinfo {author} {\bibfnamefont {T.~S.}\ \bibnamefont {Metodi}},
  \bibinfo {author} {\bibfnamefont {S.~D.}\ \bibnamefont {Gasster}},\ and\
  \bibinfo {author} {\bibfnamefont {K.~R.}\ \bibnamefont {Brown}},\ }\bibfield
  {title} {\bibinfo {title} {Resource requirements for fault-tolerant quantum
  simulation: The ground state of the transverse ising model},\ }\href
  {https://doi.org/10.1103/PhysRevA.79.062314} {\bibfield  {journal} {\bibinfo
  {journal} {Phys. Rev. A}\ }\textbf {\bibinfo {volume} {79}},\ \bibinfo
  {pages} {062314} (\bibinfo {year} {2009})}\BibitemShut {NoStop}%
\bibitem [{\citenamefont {Martinez}\ \emph {et~al.}(2016)\citenamefont
  {Martinez} \emph {et~al.}}]{Martinez:2016yna}%
  \BibitemOpen
  \bibfield  {author} {\bibinfo {author} {\bibfnamefont {E.~A.}\ \bibnamefont
  {Martinez}} \emph {et~al.},\ }\bibfield  {title} {\bibinfo {title}
  {{Real-time dynamics of lattice gauge theories with a few-qubit quantum
  computer}},\ }\href {https://doi.org/10.1038/nature18318} {\bibfield
  {journal} {\bibinfo  {journal} {Nature}\ }\textbf {\bibinfo {volume} {534}},\
  \bibinfo {pages} {516} (\bibinfo {year} {2016})},\ \Eprint
  {https://arxiv.org/abs/1605.04570} {arXiv:1605.04570 [quant-ph]} \BibitemShut
  {NoStop}%
\bibitem [{\citenamefont {You}\ \emph {et~al.}(2013)\citenamefont {You},
  \citenamefont {Geller},\ and\ \citenamefont {Stancil}}]{PhysRevA.87.032341}%
  \BibitemOpen
  \bibfield  {author} {\bibinfo {author} {\bibfnamefont {H.}~\bibnamefont
  {You}}, \bibinfo {author} {\bibfnamefont {M.~R.}\ \bibnamefont {Geller}},\
  and\ \bibinfo {author} {\bibfnamefont {P.~C.}\ \bibnamefont {Stancil}},\
  }\bibfield  {title} {\bibinfo {title} {Simulating the transverse ising model
  on a quantum computer: Error correction with the surface code},\ }\href
  {https://doi.org/10.1103/PhysRevA.87.032341} {\bibfield  {journal} {\bibinfo
  {journal} {Phys. Rev. A}\ }\textbf {\bibinfo {volume} {87}},\ \bibinfo
  {pages} {032341} (\bibinfo {year} {2013})}\BibitemShut {NoStop}%
\bibitem [{\citenamefont {Lamm}\ and\ \citenamefont
  {Lawrence}(2018{\natexlab{a}})}]{PhysRevLett.121.170501}%
  \BibitemOpen
  \bibfield  {author} {\bibinfo {author} {\bibfnamefont {H.}~\bibnamefont
  {Lamm}}\ and\ \bibinfo {author} {\bibfnamefont {S.}~\bibnamefont
  {Lawrence}},\ }\bibfield  {title} {\bibinfo {title} {Simulation of
  nonequilibrium dynamics on a quantum computer},\ }\href
  {https://doi.org/10.1103/PhysRevLett.121.170501} {\bibfield  {journal}
  {\bibinfo  {journal} {Phys. Rev. Lett.}\ }\textbf {\bibinfo {volume} {121}},\
  \bibinfo {pages} {170501} (\bibinfo {year} {2018}{\natexlab{a}})}\BibitemShut
  {NoStop}%
\bibitem [{\citenamefont
  {Cervera-Lierta}(2018)}]{CerveraLierta2018exactisingmodel}%
  \BibitemOpen
  \bibfield  {author} {\bibinfo {author} {\bibfnamefont {A.}~\bibnamefont
  {Cervera-Lierta}},\ }\bibfield  {title} {\bibinfo {title} {Exact {I}sing
  model simulation on a quantum computer},\ }\href
  {https://doi.org/10.22331/q-2018-12-21-114} {\bibfield  {journal} {\bibinfo
  {journal} {{Quantum}}\ }\textbf {\bibinfo {volume} {2}},\ \bibinfo {pages}
  {114} (\bibinfo {year} {2018})}\BibitemShut {NoStop}%
\bibitem [{\citenamefont {Barends}\ \emph {et~al.}(2016)\citenamefont
  {Barends}, \citenamefont {Shabani}, \citenamefont {Lamata}, \citenamefont
  {Kelly}, \citenamefont {Mezzacapo}, \citenamefont {Heras}, \citenamefont
  {Babbush}, \citenamefont {Fowler}, \citenamefont {Campbell}, \citenamefont
  {Chen},\ and\ \citenamefont {et~al.}}]{Barends_2016}%
  \BibitemOpen
  \bibfield  {author} {\bibinfo {author} {\bibfnamefont {R.}~\bibnamefont
  {Barends}}, \bibinfo {author} {\bibfnamefont {A.}~\bibnamefont {Shabani}},
  \bibinfo {author} {\bibfnamefont {L.}~\bibnamefont {Lamata}}, \bibinfo
  {author} {\bibfnamefont {J.}~\bibnamefont {Kelly}}, \bibinfo {author}
  {\bibfnamefont {A.}~\bibnamefont {Mezzacapo}}, \bibinfo {author}
  {\bibfnamefont {U.~L.}\ \bibnamefont {Heras}}, \bibinfo {author}
  {\bibfnamefont {R.}~\bibnamefont {Babbush}}, \bibinfo {author} {\bibfnamefont
  {A.~G.}\ \bibnamefont {Fowler}}, \bibinfo {author} {\bibfnamefont
  {B.}~\bibnamefont {Campbell}}, \bibinfo {author} {\bibfnamefont
  {Y.}~\bibnamefont {Chen}},\ and\ \bibinfo {author} {\bibnamefont {et~al.}},\
  }\bibfield  {title} {\bibinfo {title} {Digitized adiabatic quantum computing
  with a superconducting circuit},\ }\href
  {https://doi.org/10.1038/nature17658} {\bibfield  {journal} {\bibinfo
  {journal} {Nature}\ }\textbf {\bibinfo {volume} {534}},\ \bibinfo {pages}
  {222–226} (\bibinfo {year} {2016})}\BibitemShut {NoStop}%
\bibitem [{\citenamefont {Klco}\ \emph {et~al.}(2018)\citenamefont {Klco},
  \citenamefont {Dumitrescu}, \citenamefont {McCaskey}, \citenamefont {Morris},
  \citenamefont {Pooser}, \citenamefont {Sanz}, \citenamefont {Solano},
  \citenamefont {Lougovski},\ and\ \citenamefont {Savage}}]{Klco:2018kyo}%
  \BibitemOpen
  \bibfield  {author} {\bibinfo {author} {\bibfnamefont {N.}~\bibnamefont
  {Klco}}, \bibinfo {author} {\bibfnamefont {E.~F.}\ \bibnamefont
  {Dumitrescu}}, \bibinfo {author} {\bibfnamefont {A.~J.}\ \bibnamefont
  {McCaskey}}, \bibinfo {author} {\bibfnamefont {T.~D.}\ \bibnamefont
  {Morris}}, \bibinfo {author} {\bibfnamefont {R.~C.}\ \bibnamefont {Pooser}},
  \bibinfo {author} {\bibfnamefont {M.}~\bibnamefont {Sanz}}, \bibinfo {author}
  {\bibfnamefont {E.}~\bibnamefont {Solano}}, \bibinfo {author} {\bibfnamefont
  {P.}~\bibnamefont {Lougovski}},\ and\ \bibinfo {author} {\bibfnamefont
  {M.~J.}\ \bibnamefont {Savage}},\ }\bibfield  {title} {\bibinfo {title}
  {Quantum-classical computation of schwinger model dynamics using quantum
  computers},\ }\bibfield  {journal} {\bibinfo  {journal} {Physical Review A}\
  }\textbf {\bibinfo {volume} {98}},\ \href
  {https://doi.org/10.1103/physreva.98.032331} {10.1103/physreva.98.032331}
  (\bibinfo {year} {2018})\BibitemShut {NoStop}%
\bibitem [{\citenamefont {Verdel}\ \emph {et~al.}(2020)\citenamefont {Verdel},
  \citenamefont {Liu}, \citenamefont {Whitsitt}, \citenamefont {Gorshkov},\
  and\ \citenamefont {Heyl}}]{Verdel_2020}%
  \BibitemOpen
  \bibfield  {author} {\bibinfo {author} {\bibfnamefont {R.}~\bibnamefont
  {Verdel}}, \bibinfo {author} {\bibfnamefont {F.}~\bibnamefont {Liu}},
  \bibinfo {author} {\bibfnamefont {S.}~\bibnamefont {Whitsitt}}, \bibinfo
  {author} {\bibfnamefont {A.~V.}\ \bibnamefont {Gorshkov}},\ and\ \bibinfo
  {author} {\bibfnamefont {M.}~\bibnamefont {Heyl}},\ }\bibfield  {title}
  {\bibinfo {title} {Real-time dynamics of string breaking in quantum spin
  chains},\ }\bibfield  {journal} {\bibinfo  {journal} {Physical Review B}\
  }\textbf {\bibinfo {volume} {102}},\ \href
  {https://doi.org/10.1103/physrevb.102.014308} {10.1103/physrevb.102.014308}
  (\bibinfo {year} {2020})\BibitemShut {NoStop}%
\bibitem [{\citenamefont {Yamamoto}(2020)}]{Yamamoto_2020}%
  \BibitemOpen
  \bibfield  {author} {\bibinfo {author} {\bibfnamefont {A.}~\bibnamefont
  {Yamamoto}},\ }\bibfield  {title} {\bibinfo {title} {Real-time simulation of
  (2+1)-dimensional lattice gauge theory on qubits},\ }\bibfield  {journal}
  {\bibinfo  {journal} {Progress of Theoretical and Experimental Physics}\
  }\textbf {\bibinfo {volume} {2021}},\ \href
  {https://doi.org/10.1093/ptep/ptaa171} {10.1093/ptep/ptaa171} (\bibinfo
  {year} {2020})\BibitemShut {NoStop}%
\bibitem [{\citenamefont {{Borzenkova}}\ \emph {et~al.}(2020)\citenamefont
  {{Borzenkova}}, \citenamefont {{Struchalin}}, \citenamefont {{Kardashin}},
  \citenamefont {{Krasnikov}}, \citenamefont {{Skryabin}}, \citenamefont
  {{Straupe}}, \citenamefont {{Kulik}},\ and\ \citenamefont
  {{Biamonte}}}]{2020arXiv200909551B}%
  \BibitemOpen
  \bibfield  {author} {\bibinfo {author} {\bibfnamefont {O.~V.}\ \bibnamefont
  {{Borzenkova}}}, \bibinfo {author} {\bibfnamefont {G.~I.}\ \bibnamefont
  {{Struchalin}}}, \bibinfo {author} {\bibfnamefont {A.~S.}\ \bibnamefont
  {{Kardashin}}}, \bibinfo {author} {\bibfnamefont {V.~V.}\ \bibnamefont
  {{Krasnikov}}}, \bibinfo {author} {\bibfnamefont {N.~N.}\ \bibnamefont
  {{Skryabin}}}, \bibinfo {author} {\bibfnamefont {S.~S.}\ \bibnamefont
  {{Straupe}}}, \bibinfo {author} {\bibfnamefont {S.~P.}\ \bibnamefont
  {{Kulik}}},\ and\ \bibinfo {author} {\bibfnamefont {J.~D.}\ \bibnamefont
  {{Biamonte}}},\ }\bibfield  {title} {\bibinfo {title} {{Variational
  Simulation of Schwinger's Hamiltonian with Polarisation Qubits}},\
  }\href@noop {} {\bibfield  {journal} {\bibinfo  {journal} {arXiv e-prints}\
  ,\ \bibinfo {eid} {arXiv:2009.09551}} (\bibinfo {year} {2020})},\ \Eprint
  {https://arxiv.org/abs/2009.09551} {arXiv:2009.09551 [quant-ph]} \BibitemShut
  {NoStop}%
\bibitem [{\citenamefont {Notarnicola}\ \emph {et~al.}(2020)\citenamefont
  {Notarnicola}, \citenamefont {Collura},\ and\ \citenamefont
  {Montangero}}]{Notarnicola_2020}%
  \BibitemOpen
  \bibfield  {author} {\bibinfo {author} {\bibfnamefont {S.}~\bibnamefont
  {Notarnicola}}, \bibinfo {author} {\bibfnamefont {M.}~\bibnamefont
  {Collura}},\ and\ \bibinfo {author} {\bibfnamefont {S.}~\bibnamefont
  {Montangero}},\ }\bibfield  {title} {\bibinfo {title} {Real-time-dynamics
  quantum simulation of (1+1)-dimensional lattice qed with rydberg atoms},\
  }\bibfield  {journal} {\bibinfo  {journal} {Physical Review Research}\
  }\textbf {\bibinfo {volume} {2}},\ \href
  {https://doi.org/10.1103/physrevresearch.2.013288}
  {10.1103/physrevresearch.2.013288} (\bibinfo {year} {2020})\BibitemShut
  {NoStop}%
\bibitem [{\citenamefont {Brower}\ \emph
  {et~al.}(2020{\natexlab{a}})\citenamefont {Brower}, \citenamefont
  {Berenstein},\ and\ \citenamefont {Kawai}}]{Brower:2020huh}%
  \BibitemOpen
  \bibfield  {author} {\bibinfo {author} {\bibfnamefont {R.~C.}\ \bibnamefont
  {Brower}}, \bibinfo {author} {\bibfnamefont {D.}~\bibnamefont {Berenstein}},\
  and\ \bibinfo {author} {\bibfnamefont {H.}~\bibnamefont {Kawai}},\ }\bibfield
   {title} {\bibinfo {title} {{Lattice Gauge Theory for a Quantum Computer}},\
  }\href {https://doi.org/10.22323/1.363.0112} {\bibfield  {journal} {\bibinfo
  {journal} {PoS}\ }\textbf {\bibinfo {volume} {LATTICE2019}},\ \bibinfo
  {pages} {112} (\bibinfo {year} {2020}{\natexlab{a}})},\ \Eprint
  {https://arxiv.org/abs/2002.10028} {arXiv:2002.10028 [hep-lat]} \BibitemShut
  {NoStop}%
\bibitem [{\citenamefont {Lamm}\ and\ \citenamefont
  {Lawrence}(2018{\natexlab{b}})}]{Lamm:2018siq}%
  \BibitemOpen
  \bibfield  {author} {\bibinfo {author} {\bibfnamefont {H.}~\bibnamefont
  {Lamm}}\ and\ \bibinfo {author} {\bibfnamefont {S.}~\bibnamefont
  {Lawrence}},\ }\bibfield  {title} {\bibinfo {title} {{Simulation of
  Nonequilibrium Dynamics on a Quantum Computer}},\ }\href
  {https://doi.org/10.1103/PhysRevLett.121.170501} {\bibfield  {journal}
  {\bibinfo  {journal} {Phys. Rev. Lett.}\ }\textbf {\bibinfo {volume} {121}},\
  \bibinfo {pages} {170501} (\bibinfo {year} {2018}{\natexlab{b}})},\ \Eprint
  {https://arxiv.org/abs/1806.06649} {arXiv:1806.06649 [quant-ph]} \BibitemShut
  {NoStop}%
\bibitem [{\citenamefont {Gustafson}\ \emph
  {et~al.}(2019{\natexlab{a}})\citenamefont {Gustafson}, \citenamefont
  {Meurice},\ and\ \citenamefont {Unmuth-Yockey}}]{GustafsonIsing}%
  \BibitemOpen
  \bibfield  {author} {\bibinfo {author} {\bibfnamefont {E.}~\bibnamefont
  {Gustafson}}, \bibinfo {author} {\bibfnamefont {Y.}~\bibnamefont {Meurice}},\
  and\ \bibinfo {author} {\bibfnamefont {J.}~\bibnamefont {Unmuth-Yockey}},\
  }\bibfield  {title} {\bibinfo {title} {Quantum simulation of scattering in
  the quantum ising model},\ }\bibfield  {journal} {\bibinfo  {journal}
  {Physical Review D}\ }\textbf {\bibinfo {volume} {99}},\ \href
  {https://doi.org/10.1103/PhysRevD.99.094503} {10.1103/PhysRevD.99.094503}
  (\bibinfo {year} {2019}{\natexlab{a}})\BibitemShut {NoStop}%
\bibitem [{\citenamefont {Gustafson}\ \emph
  {et~al.}(2019{\natexlab{b}})\citenamefont {Gustafson}, \citenamefont
  {Dreher}, \citenamefont {Hang},\ and\ \citenamefont
  {Meurice}}]{gustafson2019real}%
  \BibitemOpen
  \bibfield  {author} {\bibinfo {author} {\bibfnamefont {E.}~\bibnamefont
  {Gustafson}}, \bibinfo {author} {\bibfnamefont {P.}~\bibnamefont {Dreher}},
  \bibinfo {author} {\bibfnamefont {Z.}~\bibnamefont {Hang}},\ and\ \bibinfo
  {author} {\bibfnamefont {Y.}~\bibnamefont {Meurice}},\ }\href@noop {}
  {\bibinfo {title} {Real time evolution of a one-dimensional field theory on a
  20 qubit machine}} (\bibinfo {year} {2019}{\natexlab{b}}),\ \Eprint
  {https://arxiv.org/abs/1910.09478} {arXiv:1910.09478 [hep-lat]} \BibitemShut
  {NoStop}%
\bibitem [{\citenamefont {Kim}\ \emph {et~al.}(2020)\citenamefont {Kim},
  \citenamefont {Song}, \citenamefont {Kim},\ and\ \citenamefont
  {Ahn}}]{Kim_2020}%
  \BibitemOpen
  \bibfield  {author} {\bibinfo {author} {\bibfnamefont {M.}~\bibnamefont
  {Kim}}, \bibinfo {author} {\bibfnamefont {Y.}~\bibnamefont {Song}}, \bibinfo
  {author} {\bibfnamefont {J.}~\bibnamefont {Kim}},\ and\ \bibinfo {author}
  {\bibfnamefont {J.}~\bibnamefont {Ahn}},\ }\bibfield  {title} {\bibinfo
  {title} {Quantum ising hamiltonian programming in trio, quartet, and sextet
  qubit systems},\ }\bibfield  {journal} {\bibinfo  {journal} {PRX Quantum}\
  }\textbf {\bibinfo {volume} {1}},\ \href
  {https://doi.org/10.1103/prxquantum.1.020323} {10.1103/prxquantum.1.020323}
  (\bibinfo {year} {2020})\BibitemShut {NoStop}%
\bibitem [{\citenamefont {Yeter-Aydeniz}\ \emph {et~al.}(2021)\citenamefont
  {Yeter-Aydeniz}, \citenamefont {Siopsis},\ and\ \citenamefont
  {Pooser}}]{yeteraydeniz2021scattering}%
  \BibitemOpen
  \bibfield  {author} {\bibinfo {author} {\bibfnamefont {K.}~\bibnamefont
  {Yeter-Aydeniz}}, \bibinfo {author} {\bibfnamefont {G.}~\bibnamefont
  {Siopsis}},\ and\ \bibinfo {author} {\bibfnamefont {R.~C.}\ \bibnamefont
  {Pooser}},\ }\href@noop {} {\bibinfo {title} {Scattering in the ising model
  using quantum lanczos algorithm}} (\bibinfo {year} {2021}),\ \Eprint
  {https://arxiv.org/abs/2008.08763} {arXiv:2008.08763 [quant-ph]} \BibitemShut
  {NoStop}%
\bibitem [{\citenamefont {Vovrosh}\ and\ \citenamefont
  {Knolle}(2020)}]{vovrosh2020confinement}%
  \BibitemOpen
  \bibfield  {author} {\bibinfo {author} {\bibfnamefont {J.}~\bibnamefont
  {Vovrosh}}\ and\ \bibinfo {author} {\bibfnamefont {J.}~\bibnamefont
  {Knolle}},\ }\bibfield  {title} {\bibinfo {title} {Confinement and
  entanglement dynamics on a digital quantum computer},\ }\href@noop {} {\
  (\bibinfo {year} {2020})},\ \Eprint {https://arxiv.org/abs/2001.03044}
  {arXiv:2001.03044 [cond-mat.str-el]} \BibitemShut {NoStop}%
\bibitem [{\citenamefont {Kandala}\ \emph {et~al.}(2017)\citenamefont
  {Kandala}, \citenamefont {Mezzacapo}, \citenamefont {Temme}, \citenamefont
  {Takita}, \citenamefont {Brink}, \citenamefont {Chow},\ and\ \citenamefont
  {Gambetta}}]{Kandala:2017aa}%
  \BibitemOpen
  \bibfield  {author} {\bibinfo {author} {\bibfnamefont {A.}~\bibnamefont
  {Kandala}}, \bibinfo {author} {\bibfnamefont {A.}~\bibnamefont {Mezzacapo}},
  \bibinfo {author} {\bibfnamefont {K.}~\bibnamefont {Temme}}, \bibinfo
  {author} {\bibfnamefont {M.}~\bibnamefont {Takita}}, \bibinfo {author}
  {\bibfnamefont {M.}~\bibnamefont {Brink}}, \bibinfo {author} {\bibfnamefont
  {J.~M.}\ \bibnamefont {Chow}},\ and\ \bibinfo {author} {\bibfnamefont
  {J.~M.}\ \bibnamefont {Gambetta}},\ }\bibfield  {title} {\bibinfo {title}
  {Hardware-efficient variational quantum eigensolver for small molecules and
  quantum magnets},\ }\href {https://doi.org/10.1038/nature23879} {\bibfield
  {journal} {\bibinfo  {journal} {Nature}\ }\textbf {\bibinfo {volume} {549}},\
  \bibinfo {pages} {242 EP } (\bibinfo {year} {2017})}\BibitemShut {NoStop}%
\bibitem [{\citenamefont {Kandala}\ \emph {et~al.}(2019)\citenamefont
  {Kandala}, \citenamefont {Temme}, \citenamefont {Córcoles}, \citenamefont
  {Mezzacapo}, \citenamefont {Chow},\ and\ \citenamefont
  {Gambetta}}]{Kandala_2019}%
  \BibitemOpen
  \bibfield  {author} {\bibinfo {author} {\bibfnamefont {A.}~\bibnamefont
  {Kandala}}, \bibinfo {author} {\bibfnamefont {K.}~\bibnamefont {Temme}},
  \bibinfo {author} {\bibfnamefont {A.~D.}\ \bibnamefont {Córcoles}}, \bibinfo
  {author} {\bibfnamefont {A.}~\bibnamefont {Mezzacapo}}, \bibinfo {author}
  {\bibfnamefont {J.~M.}\ \bibnamefont {Chow}},\ and\ \bibinfo {author}
  {\bibfnamefont {J.~M.}\ \bibnamefont {Gambetta}},\ }\bibfield  {title}
  {\bibinfo {title} {Error mitigation extends the computational reach of a
  noisy quantum processor},\ }\href {https://doi.org/10.1038/s41586-019-1040-7}
  {\bibfield  {journal} {\bibinfo  {journal} {Nature}\ }\textbf {\bibinfo
  {volume} {567}},\ \bibinfo {pages} {491–495} (\bibinfo {year}
  {2019})}\BibitemShut {NoStop}%
\bibitem [{\citenamefont {Salathé}\ \emph {et~al.}(2015)\citenamefont
  {Salathé}, \citenamefont {Mondal}, \citenamefont {Oppliger}, \citenamefont
  {Heinsoo}, \citenamefont {Kurpiers}, \citenamefont {Potočnik}, \citenamefont
  {Mezzacapo}, \citenamefont {Las~Heras}, \citenamefont {Lamata}, \citenamefont
  {Solano},\ and\ \citenamefont {et~al.}}]{Salath__2015}%
  \BibitemOpen
  \bibfield  {author} {\bibinfo {author} {\bibfnamefont {Y.}~\bibnamefont
  {Salathé}}, \bibinfo {author} {\bibfnamefont {M.}~\bibnamefont {Mondal}},
  \bibinfo {author} {\bibfnamefont {M.}~\bibnamefont {Oppliger}}, \bibinfo
  {author} {\bibfnamefont {J.}~\bibnamefont {Heinsoo}}, \bibinfo {author}
  {\bibfnamefont {P.}~\bibnamefont {Kurpiers}}, \bibinfo {author}
  {\bibfnamefont {A.}~\bibnamefont {Potočnik}}, \bibinfo {author}
  {\bibfnamefont {A.}~\bibnamefont {Mezzacapo}}, \bibinfo {author}
  {\bibfnamefont {U.}~\bibnamefont {Las~Heras}}, \bibinfo {author}
  {\bibfnamefont {L.}~\bibnamefont {Lamata}}, \bibinfo {author} {\bibfnamefont
  {E.}~\bibnamefont {Solano}},\ and\ \bibinfo {author} {\bibnamefont
  {et~al.}},\ }\bibfield  {title} {\bibinfo {title} {Digital quantum simulation
  of spin models with circuit quantum electrodynamics},\ }\bibfield  {journal}
  {\bibinfo  {journal} {Physical Review X}\ }\textbf {\bibinfo {volume} {5}},\
  \href {https://doi.org/10.1103/physrevx.5.021027} {10.1103/physrevx.5.021027}
  (\bibinfo {year} {2015})\BibitemShut {NoStop}%
\bibitem [{\citenamefont {Labuhn}\ \emph {et~al.}(2016)\citenamefont {Labuhn},
  \citenamefont {Barredo}, \citenamefont {Ravets}, \citenamefont
  {de~Léséleuc}, \citenamefont {Macrì}, \citenamefont {Lahaye},\ and\
  \citenamefont {Browaeys}}]{Labuhn_2016}%
  \BibitemOpen
  \bibfield  {author} {\bibinfo {author} {\bibfnamefont {H.}~\bibnamefont
  {Labuhn}}, \bibinfo {author} {\bibfnamefont {D.}~\bibnamefont {Barredo}},
  \bibinfo {author} {\bibfnamefont {S.}~\bibnamefont {Ravets}}, \bibinfo
  {author} {\bibfnamefont {S.}~\bibnamefont {de~Léséleuc}}, \bibinfo {author}
  {\bibfnamefont {T.}~\bibnamefont {Macrì}}, \bibinfo {author} {\bibfnamefont
  {T.}~\bibnamefont {Lahaye}},\ and\ \bibinfo {author} {\bibfnamefont
  {A.}~\bibnamefont {Browaeys}},\ }\bibfield  {title} {\bibinfo {title}
  {Tunable two-dimensional arrays of single rydberg atoms for realizing quantum
  ising models},\ }\href {https://doi.org/10.1038/nature18274} {\bibfield
  {journal} {\bibinfo  {journal} {Nature}\ }\textbf {\bibinfo {volume} {534}},\
  \bibinfo {pages} {667–670} (\bibinfo {year} {2016})}\BibitemShut {NoStop}%
\bibitem [{\citenamefont {{Zhang}}\ \emph {et~al.}(2017)\citenamefont
  {{Zhang}}, \citenamefont {{Pagano}}, \citenamefont {{Hess}}, \citenamefont
  {{Kyprianidis}}, \citenamefont {{Becker}}, \citenamefont {{Kaplan}},
  \citenamefont {{Gorshkov}}, \citenamefont {{Gong}},\ and\ \citenamefont
  {{Monroe}}}]{2017Natur.551..601Z}%
  \BibitemOpen
  \bibfield  {author} {\bibinfo {author} {\bibfnamefont {J.}~\bibnamefont
  {{Zhang}}}, \bibinfo {author} {\bibfnamefont {G.}~\bibnamefont {{Pagano}}},
  \bibinfo {author} {\bibfnamefont {P.~W.}\ \bibnamefont {{Hess}}}, \bibinfo
  {author} {\bibfnamefont {A.}~\bibnamefont {{Kyprianidis}}}, \bibinfo {author}
  {\bibfnamefont {P.}~\bibnamefont {{Becker}}}, \bibinfo {author}
  {\bibfnamefont {H.}~\bibnamefont {{Kaplan}}}, \bibinfo {author}
  {\bibfnamefont {A.~V.}\ \bibnamefont {{Gorshkov}}}, \bibinfo {author}
  {\bibfnamefont {Z.~X.}\ \bibnamefont {{Gong}}},\ and\ \bibinfo {author}
  {\bibfnamefont {C.}~\bibnamefont {{Monroe}}},\ }\bibfield  {title} {\bibinfo
  {title} {{Observation of a many-body dynamical phase transition with a
  53-qubit quantum simulator}},\ }\href {https://doi.org/10.1038/nature24654}
  {\bibfield  {journal} {\bibinfo  {journal} {\nat}\ }\textbf {\bibinfo
  {volume} {551}},\ \bibinfo {pages} {601} (\bibinfo {year} {2017})},\ \Eprint
  {https://arxiv.org/abs/1708.01044} {arXiv:1708.01044 [quant-ph]} \BibitemShut
  {NoStop}%
\bibitem [{\citenamefont {Kadowaki}\ and\ \citenamefont
  {Nishimori}(1998)}]{PhysRevE.58.5355}%
  \BibitemOpen
  \bibfield  {author} {\bibinfo {author} {\bibfnamefont {T.}~\bibnamefont
  {Kadowaki}}\ and\ \bibinfo {author} {\bibfnamefont {H.}~\bibnamefont
  {Nishimori}},\ }\bibfield  {title} {\bibinfo {title} {Quantum annealing in
  the transverse ising model},\ }\href
  {https://doi.org/10.1103/PhysRevE.58.5355} {\bibfield  {journal} {\bibinfo
  {journal} {Phys. Rev. E}\ }\textbf {\bibinfo {volume} {58}},\ \bibinfo
  {pages} {5355} (\bibinfo {year} {1998})}\BibitemShut {NoStop}%
\bibitem [{\citenamefont {{Bernien}}\ \emph {et~al.}(2017)\citenamefont
  {{Bernien}}, \citenamefont {{Schwartz}}, \citenamefont {{Keesling}},
  \citenamefont {{Levine}}, \citenamefont {{Omran}}, \citenamefont {{Pichler}},
  \citenamefont {{Choi}}, \citenamefont {{Zibrov}}, \citenamefont {{Endres}},
  \citenamefont {{Greiner}}, \citenamefont {{Vuleti{\'c}}},\ and\ \citenamefont
  {{Lukin}}}]{2017Natur.551..579B}%
  \BibitemOpen
  \bibfield  {author} {\bibinfo {author} {\bibfnamefont {H.}~\bibnamefont
  {{Bernien}}}, \bibinfo {author} {\bibfnamefont {S.}~\bibnamefont
  {{Schwartz}}}, \bibinfo {author} {\bibfnamefont {A.}~\bibnamefont
  {{Keesling}}}, \bibinfo {author} {\bibfnamefont {H.}~\bibnamefont
  {{Levine}}}, \bibinfo {author} {\bibfnamefont {A.}~\bibnamefont {{Omran}}},
  \bibinfo {author} {\bibfnamefont {H.}~\bibnamefont {{Pichler}}}, \bibinfo
  {author} {\bibfnamefont {S.}~\bibnamefont {{Choi}}}, \bibinfo {author}
  {\bibfnamefont {A.~S.}\ \bibnamefont {{Zibrov}}}, \bibinfo {author}
  {\bibfnamefont {M.}~\bibnamefont {{Endres}}}, \bibinfo {author}
  {\bibfnamefont {M.}~\bibnamefont {{Greiner}}}, \bibinfo {author}
  {\bibfnamefont {V.}~\bibnamefont {{Vuleti{\'c}}}},\ and\ \bibinfo {author}
  {\bibfnamefont {M.~D.}\ \bibnamefont {{Lukin}}},\ }\bibfield  {title}
  {\bibinfo {title} {{Probing many-body dynamics on a 51-atom quantum
  simulator}},\ }\href {https://doi.org/10.1038/nature24622} {\bibfield
  {journal} {\bibinfo  {journal} {\nat}\ }\textbf {\bibinfo {volume} {551}},\
  \bibinfo {pages} {579} (\bibinfo {year} {2017})},\ \Eprint
  {https://arxiv.org/abs/1707.04344} {arXiv:1707.04344 [quant-ph]} \BibitemShut
  {NoStop}%
\bibitem [{\citenamefont {Brower}\ \emph
  {et~al.}(2020{\natexlab{b}})\citenamefont {Brower}, \citenamefont
  {Berenstein},\ and\ \citenamefont {Kawai}}]{brower2020lattice}%
  \BibitemOpen
  \bibfield  {author} {\bibinfo {author} {\bibfnamefont {R.~C.}\ \bibnamefont
  {Brower}}, \bibinfo {author} {\bibfnamefont {D.}~\bibnamefont {Berenstein}},\
  and\ \bibinfo {author} {\bibfnamefont {H.}~\bibnamefont {Kawai}},\
  }\href@noop {} {\bibinfo {title} {Lattice gauge theory for a quantum
  computer}} (\bibinfo {year} {2020}{\natexlab{b}}),\ \Eprint
  {https://arxiv.org/abs/2002.10028} {arXiv:2002.10028 [hep-lat]} \BibitemShut
  {NoStop}%
\bibitem [{\citenamefont {Choi}\ \emph {et~al.}(2017)\citenamefont {Choi},
  \citenamefont {Yao},\ and\ \citenamefont {Lukin}}]{Choi_2017}%
  \BibitemOpen
  \bibfield  {author} {\bibinfo {author} {\bibfnamefont {S.}~\bibnamefont
  {Choi}}, \bibinfo {author} {\bibfnamefont {N.~Y.}\ \bibnamefont {Yao}},\ and\
  \bibinfo {author} {\bibfnamefont {M.~D.}\ \bibnamefont {Lukin}},\ }\bibfield
  {title} {\bibinfo {title} {Dynamical engineering of interactions in qudit
  ensembles},\ }\bibfield  {journal} {\bibinfo  {journal} {Physical Review
  Letters}\ }\textbf {\bibinfo {volume} {119}},\ \href
  {https://doi.org/10.1103/physrevlett.119.183603}
  {10.1103/physrevlett.119.183603} (\bibinfo {year} {2017})\BibitemShut
  {NoStop}%
\bibitem [{\citenamefont {Bender}\ \emph {et~al.}(2020)\citenamefont {Bender},
  \citenamefont {Emonts}, \citenamefont {Zohar},\ and\ \citenamefont
  {Cirac}}]{Bender_2020}%
  \BibitemOpen
  \bibfield  {author} {\bibinfo {author} {\bibfnamefont {J.}~\bibnamefont
  {Bender}}, \bibinfo {author} {\bibfnamefont {P.}~\bibnamefont {Emonts}},
  \bibinfo {author} {\bibfnamefont {E.}~\bibnamefont {Zohar}},\ and\ \bibinfo
  {author} {\bibfnamefont {J.~I.}\ \bibnamefont {Cirac}},\ }\bibfield  {title}
  {\bibinfo {title} {Real-time dynamics in 2+1d compact qed using complex
  periodic gaussian states},\ }\bibfield  {journal} {\bibinfo  {journal}
  {Physical Review Research}\ }\textbf {\bibinfo {volume} {2}},\ \href
  {https://doi.org/10.1103/physrevresearch.2.043145}
  {10.1103/physrevresearch.2.043145} (\bibinfo {year} {2020})\BibitemShut
  {NoStop}%
\bibitem [{\citenamefont {Zohar}\ \emph {et~al.}(2012)\citenamefont {Zohar},
  \citenamefont {Cirac},\ and\ \citenamefont {Reznik}}]{Zohar_2012}%
  \BibitemOpen
  \bibfield  {author} {\bibinfo {author} {\bibfnamefont {E.}~\bibnamefont
  {Zohar}}, \bibinfo {author} {\bibfnamefont {J.~I.}\ \bibnamefont {Cirac}},\
  and\ \bibinfo {author} {\bibfnamefont {B.}~\bibnamefont {Reznik}},\
  }\bibfield  {title} {\bibinfo {title} {Simulating compact quantum
  electrodynamics with ultracold atoms: Probing confinement and nonperturbative
  effects},\ }\bibfield  {journal} {\bibinfo  {journal} {Physical Review
  Letters}\ }\textbf {\bibinfo {volume} {109}},\ \href
  {https://doi.org/10.1103/physrevlett.109.125302}
  {10.1103/physrevlett.109.125302} (\bibinfo {year} {2012})\BibitemShut
  {NoStop}%
\bibitem [{\citenamefont {Di}\ and\ \citenamefont
  {Wei}(2012)}]{di2012elementary}%
  \BibitemOpen
  \bibfield  {author} {\bibinfo {author} {\bibfnamefont {Y.-M.}\ \bibnamefont
  {Di}}\ and\ \bibinfo {author} {\bibfnamefont {H.-R.}\ \bibnamefont {Wei}},\
  }\bibfield  {title} {\bibinfo {title} {Elementary gates for ternary quantum
  logic circuit},\ }\href@noop {} {\  (\bibinfo {year} {2012})},\ \Eprint
  {https://arxiv.org/abs/1105.5485} {arXiv:1105.5485 [quant-ph]} \BibitemShut
  {NoStop}%
\bibitem [{\citenamefont {Baker}\ \emph {et~al.}(2020)\citenamefont {Baker},
  \citenamefont {Duckering},\ and\ \citenamefont {Chong}}]{baker2020efficient}%
  \BibitemOpen
  \bibfield  {author} {\bibinfo {author} {\bibfnamefont {J.~M.}\ \bibnamefont
  {Baker}}, \bibinfo {author} {\bibfnamefont {C.}~\bibnamefont {Duckering}},\
  and\ \bibinfo {author} {\bibfnamefont {F.~T.}\ \bibnamefont {Chong}},\
  }\bibfield  {title} {\bibinfo {title} {Efficient quantum circuit
  decompositions via intermediate qudits},\ }\href@noop {} {\  (\bibinfo {year}
  {2020})},\ \Eprint {https://arxiv.org/abs/2002.10592} {arXiv:2002.10592
  [quant-ph]} \BibitemShut {NoStop}%
\bibitem [{\citenamefont {{Ralph}}\ \emph {et~al.}(2007)\citenamefont
  {{Ralph}}, \citenamefont {{Resch}},\ and\ \citenamefont
  {{Gilchrist}}}]{2007PhRvA.75b2313R}%
  \BibitemOpen
  \bibfield  {author} {\bibinfo {author} {\bibfnamefont {T.~C.}\ \bibnamefont
  {{Ralph}}}, \bibinfo {author} {\bibfnamefont {K.~J.}\ \bibnamefont
  {{Resch}}},\ and\ \bibinfo {author} {\bibfnamefont {A.}~\bibnamefont
  {{Gilchrist}}},\ }\bibfield  {title} {\bibinfo {title} {{Efficient Toffoli
  gates using qudits}},\ }\href {https://doi.org/10.1103/PhysRevA.75.022313}
  {\bibfield  {journal} {\bibinfo  {journal} {\pra}\ }\textbf {\bibinfo
  {volume} {75}},\ \bibinfo {eid} {022313} (\bibinfo {year} {2007})},\ \Eprint
  {https://arxiv.org/abs/0806.0654} {arXiv:0806.0654 [quant-ph]} \BibitemShut
  {NoStop}%
\bibitem [{\citenamefont {{Gedik}}\ \emph {et~al.}(2015)\citenamefont
  {{Gedik}}, \citenamefont {{Silva}}, \citenamefont {{{\c{C}}akmak}},
  \citenamefont {{Karpat}}, \citenamefont {{Vidoto}}, \citenamefont
  {{Soares-Pinto}}, \citenamefont {{Deazevedo}},\ and\ \citenamefont
  {{Fanchini}}}]{2015NatSR.514671G}%
  \BibitemOpen
  \bibfield  {author} {\bibinfo {author} {\bibfnamefont {Z.}~\bibnamefont
  {{Gedik}}}, \bibinfo {author} {\bibfnamefont {I.~A.}\ \bibnamefont
  {{Silva}}}, \bibinfo {author} {\bibfnamefont {B.}~\bibnamefont
  {{{\c{C}}akmak}}}, \bibinfo {author} {\bibfnamefont {G.}~\bibnamefont
  {{Karpat}}}, \bibinfo {author} {\bibfnamefont {E.~L.~G.}\ \bibnamefont
  {{Vidoto}}}, \bibinfo {author} {\bibfnamefont {D.~O.}\ \bibnamefont
  {{Soares-Pinto}}}, \bibinfo {author} {\bibfnamefont {E.~R.}\ \bibnamefont
  {{Deazevedo}}},\ and\ \bibinfo {author} {\bibfnamefont {F.~F.}\ \bibnamefont
  {{Fanchini}}},\ }\bibfield  {title} {\bibinfo {title} {{Computational
  speed-up with a single qudit}},\ }\href {https://doi.org/10.1038/srep14671}
  {\bibfield  {journal} {\bibinfo  {journal} {Scientific Reports}\ }\textbf
  {\bibinfo {volume} {5}},\ \bibinfo {eid} {14671} (\bibinfo {year} {2015})},\
  \Eprint {https://arxiv.org/abs/1403.5861} {arXiv:1403.5861 [quant-ph]}
  \BibitemShut {NoStop}%
\bibitem [{\citenamefont {Napolitano}\ \emph {et~al.}(2021)\citenamefont
  {Napolitano}, \citenamefont {Fanchini}, \citenamefont {da~Silva},\ and\
  \citenamefont {Bellomo}}]{Napolitano_2021}%
  \BibitemOpen
  \bibfield  {author} {\bibinfo {author} {\bibfnamefont {R.~d.~J.}\
  \bibnamefont {Napolitano}}, \bibinfo {author} {\bibfnamefont {F.~F.}\
  \bibnamefont {Fanchini}}, \bibinfo {author} {\bibfnamefont {A.~H.}\
  \bibnamefont {da~Silva}},\ and\ \bibinfo {author} {\bibfnamefont
  {B.}~\bibnamefont {Bellomo}},\ }\bibfield  {title} {\bibinfo {title}
  {Protecting operations on qudits from noise by continuous dynamical
  decoupling},\ }\bibfield  {journal} {\bibinfo  {journal} {Physical Review
  Research}\ }\textbf {\bibinfo {volume} {3}},\ \href
  {https://doi.org/10.1103/physrevresearch.3.013235}
  {10.1103/physrevresearch.3.013235} (\bibinfo {year} {2021})\BibitemShut
  {NoStop}%
\bibitem [{\citenamefont {Gedik}\ \emph {et~al.}(2015)\citenamefont {Gedik},
  \citenamefont {Silva}, \citenamefont {Çakmak}, \citenamefont {Karpat},
  \citenamefont {Vidoto}, \citenamefont {Soares-Pinto}, \citenamefont
  {deAzevedo},\ and\ \citenamefont {Fanchini}}]{Gedik_2015}%
  \BibitemOpen
  \bibfield  {author} {\bibinfo {author} {\bibfnamefont {Z.}~\bibnamefont
  {Gedik}}, \bibinfo {author} {\bibfnamefont {I.~A.}\ \bibnamefont {Silva}},
  \bibinfo {author} {\bibfnamefont {B.}~\bibnamefont {Çakmak}}, \bibinfo
  {author} {\bibfnamefont {G.}~\bibnamefont {Karpat}}, \bibinfo {author}
  {\bibfnamefont {E.~L.~G.}\ \bibnamefont {Vidoto}}, \bibinfo {author}
  {\bibfnamefont {D.~O.}\ \bibnamefont {Soares-Pinto}}, \bibinfo {author}
  {\bibfnamefont {E.~R.}\ \bibnamefont {deAzevedo}},\ and\ \bibinfo {author}
  {\bibfnamefont {F.~F.}\ \bibnamefont {Fanchini}},\ }\bibfield  {title}
  {\bibinfo {title} {Computational speed-up with a single qudit},\ }\bibfield
  {journal} {\bibinfo  {journal} {Scientific Reports}\ }\textbf {\bibinfo
  {volume} {5}},\ \href {https://doi.org/10.1038/srep14671} {10.1038/srep14671}
  (\bibinfo {year} {2015})\BibitemShut {NoStop}%
\bibitem [{\citenamefont {Luo}\ \emph {et~al.}(2019)\citenamefont {Luo},
  \citenamefont {Zhong}, \citenamefont {Erhard}, \citenamefont {Wang},
  \citenamefont {Peng}, \citenamefont {Krenn}, \citenamefont {Jiang},
  \citenamefont {Li}, \citenamefont {Liu}, \citenamefont {Lu}, \citenamefont
  {Zeilinger},\ and\ \citenamefont {Pan}}]{PhysRevLett.123.070505}%
  \BibitemOpen
  \bibfield  {author} {\bibinfo {author} {\bibfnamefont {Y.-H.}\ \bibnamefont
  {Luo}}, \bibinfo {author} {\bibfnamefont {H.-S.}\ \bibnamefont {Zhong}},
  \bibinfo {author} {\bibfnamefont {M.}~\bibnamefont {Erhard}}, \bibinfo
  {author} {\bibfnamefont {X.-L.}\ \bibnamefont {Wang}}, \bibinfo {author}
  {\bibfnamefont {L.-C.}\ \bibnamefont {Peng}}, \bibinfo {author}
  {\bibfnamefont {M.}~\bibnamefont {Krenn}}, \bibinfo {author} {\bibfnamefont
  {X.}~\bibnamefont {Jiang}}, \bibinfo {author} {\bibfnamefont
  {L.}~\bibnamefont {Li}}, \bibinfo {author} {\bibfnamefont {N.-L.}\
  \bibnamefont {Liu}}, \bibinfo {author} {\bibfnamefont {C.-Y.}\ \bibnamefont
  {Lu}}, \bibinfo {author} {\bibfnamefont {A.}~\bibnamefont {Zeilinger}},\ and\
  \bibinfo {author} {\bibfnamefont {J.-W.}\ \bibnamefont {Pan}},\ }\bibfield
  {title} {\bibinfo {title} {Quantum teleportation in high dimensions},\ }\href
  {https://doi.org/10.1103/PhysRevLett.123.070505} {\bibfield  {journal}
  {\bibinfo  {journal} {Phys. Rev. Lett.}\ }\textbf {\bibinfo {volume} {123}},\
  \bibinfo {pages} {070505} (\bibinfo {year} {2019})}\BibitemShut {NoStop}%
\bibitem [{\citenamefont {Gokhale}\ \emph {et~al.}(2019)\citenamefont
  {Gokhale}, \citenamefont {Baker}, \citenamefont {Duckering}, \citenamefont
  {Brown}, \citenamefont {Brown},\ and\ \citenamefont {Chong}}]{Gokhale_2019}%
  \BibitemOpen
  \bibfield  {author} {\bibinfo {author} {\bibfnamefont {P.}~\bibnamefont
  {Gokhale}}, \bibinfo {author} {\bibfnamefont {J.~M.}\ \bibnamefont {Baker}},
  \bibinfo {author} {\bibfnamefont {C.}~\bibnamefont {Duckering}}, \bibinfo
  {author} {\bibfnamefont {N.~C.}\ \bibnamefont {Brown}}, \bibinfo {author}
  {\bibfnamefont {K.~R.}\ \bibnamefont {Brown}},\ and\ \bibinfo {author}
  {\bibfnamefont {F.~T.}\ \bibnamefont {Chong}},\ }\bibfield  {title} {\bibinfo
  {title} {Asymptotic improvements to quantum circuits via qutrits},\
  }\bibfield  {journal} {\bibinfo  {journal} {Proceedings of the 46th
  International Symposium on Computer Architecture}\ }\href
  {https://doi.org/10.1145/3307650.3322253} {10.1145/3307650.3322253} (\bibinfo
  {year} {2019})\BibitemShut {NoStop}%
\bibitem [{\citenamefont {Lapkiewicz}\ \emph {et~al.}(2011)\citenamefont
  {Lapkiewicz}, \citenamefont {Li}, \citenamefont {Schaeff}, \citenamefont
  {Langford}, \citenamefont {Ramelow}, \citenamefont {Wieśniak},\ and\
  \citenamefont {Zeilinger}}]{Lapkiewicz_2011}%
  \BibitemOpen
  \bibfield  {author} {\bibinfo {author} {\bibfnamefont {R.}~\bibnamefont
  {Lapkiewicz}}, \bibinfo {author} {\bibfnamefont {P.}~\bibnamefont {Li}},
  \bibinfo {author} {\bibfnamefont {C.}~\bibnamefont {Schaeff}}, \bibinfo
  {author} {\bibfnamefont {N.~K.}\ \bibnamefont {Langford}}, \bibinfo {author}
  {\bibfnamefont {S.}~\bibnamefont {Ramelow}}, \bibinfo {author} {\bibfnamefont
  {M.}~\bibnamefont {Wieśniak}},\ and\ \bibinfo {author} {\bibfnamefont
  {A.}~\bibnamefont {Zeilinger}},\ }\bibfield  {title} {\bibinfo {title}
  {Experimental non-classicality of an indivisible quantum system},\ }\href
  {https://doi.org/10.1038/nature10119} {\bibfield  {journal} {\bibinfo
  {journal} {Nature}\ }\textbf {\bibinfo {volume} {474}},\ \bibinfo {pages}
  {490–493} (\bibinfo {year} {2011})}\BibitemShut {NoStop}%
\bibitem [{\citenamefont {Yurtalan}\ \emph {et~al.}(2020)\citenamefont
  {Yurtalan}, \citenamefont {Shi}, \citenamefont {Kononenko}, \citenamefont
  {Lupascu},\ and\ \citenamefont {Ashhab}}]{Yurtalan_2020}%
  \BibitemOpen
  \bibfield  {author} {\bibinfo {author} {\bibfnamefont {M.}~\bibnamefont
  {Yurtalan}}, \bibinfo {author} {\bibfnamefont {J.}~\bibnamefont {Shi}},
  \bibinfo {author} {\bibfnamefont {M.}~\bibnamefont {Kononenko}}, \bibinfo
  {author} {\bibfnamefont {A.}~\bibnamefont {Lupascu}},\ and\ \bibinfo {author}
  {\bibfnamefont {S.}~\bibnamefont {Ashhab}},\ }\bibfield  {title} {\bibinfo
  {title} {Implementation of a walsh-hadamard gate in a superconducting
  qutrit},\ }\bibfield  {journal} {\bibinfo  {journal} {Physical Review
  Letters}\ }\textbf {\bibinfo {volume} {125}},\ \href
  {https://doi.org/10.1103/physrevlett.125.180504}
  {10.1103/physrevlett.125.180504} (\bibinfo {year} {2020})\BibitemShut
  {NoStop}%
\bibitem [{\citenamefont {Kononenko}\ \emph {et~al.}(2020)\citenamefont
  {Kononenko}, \citenamefont {Yurtalan}, \citenamefont {Shi},\ and\
  \citenamefont {Lupascu}}]{kononenko2020characterization}%
  \BibitemOpen
  \bibfield  {author} {\bibinfo {author} {\bibfnamefont {M.}~\bibnamefont
  {Kononenko}}, \bibinfo {author} {\bibfnamefont {M.~A.}\ \bibnamefont
  {Yurtalan}}, \bibinfo {author} {\bibfnamefont {J.}~\bibnamefont {Shi}},\ and\
  \bibinfo {author} {\bibfnamefont {A.}~\bibnamefont {Lupascu}},\ }\href@noop
  {} {\bibinfo {title} {Characterization of control in a superconducting qutrit
  using randomized benchmarking}} (\bibinfo {year} {2020}),\ \Eprint
  {https://arxiv.org/abs/2009.00599} {arXiv:2009.00599 [quant-ph]} \BibitemShut
  {NoStop}%
\bibitem [{\citenamefont {{Bianchetti}}\ \emph {et~al.}(2010)\citenamefont
  {{Bianchetti}}, \citenamefont {{Filipp}}, \citenamefont {{Baur}},
  \citenamefont {{Fink}}, \citenamefont {{Lang}}, \citenamefont {{Steffen}},
  \citenamefont {{Boissonneault}}, \citenamefont {{Blais}},\ and\ \citenamefont
  {{Wallraff}}}]{2010PhRvL.105v3601B}%
  \BibitemOpen
  \bibfield  {author} {\bibinfo {author} {\bibfnamefont {R.}~\bibnamefont
  {{Bianchetti}}}, \bibinfo {author} {\bibfnamefont {S.}~\bibnamefont
  {{Filipp}}}, \bibinfo {author} {\bibfnamefont {M.}~\bibnamefont {{Baur}}},
  \bibinfo {author} {\bibfnamefont {J.~M.}\ \bibnamefont {{Fink}}}, \bibinfo
  {author} {\bibfnamefont {C.}~\bibnamefont {{Lang}}}, \bibinfo {author}
  {\bibfnamefont {L.}~\bibnamefont {{Steffen}}}, \bibinfo {author}
  {\bibfnamefont {M.}~\bibnamefont {{Boissonneault}}}, \bibinfo {author}
  {\bibfnamefont {A.}~\bibnamefont {{Blais}}},\ and\ \bibinfo {author}
  {\bibfnamefont {A.}~\bibnamefont {{Wallraff}}},\ }\bibfield  {title}
  {\bibinfo {title} {{Control and Tomography of a Three Level Superconducting
  Artificial Atom}},\ }\href {https://doi.org/10.1103/PhysRevLett.105.223601}
  {\bibfield  {journal} {\bibinfo  {journal} {\prl}\ }\textbf {\bibinfo
  {volume} {105}},\ \bibinfo {eid} {223601} (\bibinfo {year} {2010})},\ \Eprint
  {https://arxiv.org/abs/1004.5504} {arXiv:1004.5504 [quant-ph]} \BibitemShut
  {NoStop}%
\bibitem [{\citenamefont {Lanyon}\ \emph {et~al.}(2008)\citenamefont {Lanyon},
  \citenamefont {Weinhold}, \citenamefont {Langford}, \citenamefont {O'Brien},
  \citenamefont {Resch}, \citenamefont {Gilchrist},\ and\ \citenamefont
  {White}}]{PhysRevLett.100.060504}%
  \BibitemOpen
  \bibfield  {author} {\bibinfo {author} {\bibfnamefont {B.~P.}\ \bibnamefont
  {Lanyon}}, \bibinfo {author} {\bibfnamefont {T.~J.}\ \bibnamefont
  {Weinhold}}, \bibinfo {author} {\bibfnamefont {N.~K.}\ \bibnamefont
  {Langford}}, \bibinfo {author} {\bibfnamefont {J.~L.}\ \bibnamefont
  {O'Brien}}, \bibinfo {author} {\bibfnamefont {K.~J.}\ \bibnamefont {Resch}},
  \bibinfo {author} {\bibfnamefont {A.}~\bibnamefont {Gilchrist}},\ and\
  \bibinfo {author} {\bibfnamefont {A.~G.}\ \bibnamefont {White}},\ }\bibfield
  {title} {\bibinfo {title} {Manipulating biphotonic qutrits},\ }\href
  {https://doi.org/10.1103/PhysRevLett.100.060504} {\bibfield  {journal}
  {\bibinfo  {journal} {Phys. Rev. Lett.}\ }\textbf {\bibinfo {volume} {100}},\
  \bibinfo {pages} {060504} (\bibinfo {year} {2008})}\BibitemShut {NoStop}%
\bibitem [{\citenamefont {Blok}\ \emph {et~al.}(2020)\citenamefont {Blok},
  \citenamefont {Ramasesh}, \citenamefont {Schuster}, \citenamefont {O'Brien},
  \citenamefont {Kreikebaum}, \citenamefont {Dahlen}, \citenamefont {Morvan},
  \citenamefont {Yoshida}, \citenamefont {Yao},\ and\ \citenamefont
  {Siddiqi}}]{Blok:2020may}%
  \BibitemOpen
  \bibfield  {author} {\bibinfo {author} {\bibfnamefont {M.~S.}\ \bibnamefont
  {Blok}}, \bibinfo {author} {\bibfnamefont {V.~V.}\ \bibnamefont {Ramasesh}},
  \bibinfo {author} {\bibfnamefont {T.}~\bibnamefont {Schuster}}, \bibinfo
  {author} {\bibfnamefont {K.}~\bibnamefont {O'Brien}}, \bibinfo {author}
  {\bibfnamefont {J.~M.}\ \bibnamefont {Kreikebaum}}, \bibinfo {author}
  {\bibfnamefont {D.}~\bibnamefont {Dahlen}}, \bibinfo {author} {\bibfnamefont
  {A.}~\bibnamefont {Morvan}}, \bibinfo {author} {\bibfnamefont
  {B.}~\bibnamefont {Yoshida}}, \bibinfo {author} {\bibfnamefont {N.~Y.}\
  \bibnamefont {Yao}},\ and\ \bibinfo {author} {\bibfnamefont {I.}~\bibnamefont
  {Siddiqi}},\ }\bibfield  {title} {\bibinfo {title} {{Quantum Information
  Scrambling in a Superconducting Qutrit Processor}},\ }\href@noop {} {\
  (\bibinfo {year} {2020})},\ \Eprint {https://arxiv.org/abs/2003.03307}
  {arXiv:2003.03307 [quant-ph]} \BibitemShut {NoStop}%
\bibitem [{\citenamefont {Zhang}\ \emph {et~al.}(2019)\citenamefont {Zhang},
  \citenamefont {Zhao}, \citenamefont {Wang}, \citenamefont {Xiang},
  \citenamefont {Jia}, \citenamefont {Duan}, \citenamefont {Tong},
  \citenamefont {Yin},\ and\ \citenamefont {Guo}}]{Zhang_2019}%
  \BibitemOpen
  \bibfield  {author} {\bibinfo {author} {\bibfnamefont {Z.}~\bibnamefont
  {Zhang}}, \bibinfo {author} {\bibfnamefont {P.~Z.}\ \bibnamefont {Zhao}},
  \bibinfo {author} {\bibfnamefont {T.}~\bibnamefont {Wang}}, \bibinfo {author}
  {\bibfnamefont {L.}~\bibnamefont {Xiang}}, \bibinfo {author} {\bibfnamefont
  {Z.}~\bibnamefont {Jia}}, \bibinfo {author} {\bibfnamefont {P.}~\bibnamefont
  {Duan}}, \bibinfo {author} {\bibfnamefont {D.~M.}\ \bibnamefont {Tong}},
  \bibinfo {author} {\bibfnamefont {Y.}~\bibnamefont {Yin}},\ and\ \bibinfo
  {author} {\bibfnamefont {G.}~\bibnamefont {Guo}},\ }\bibfield  {title}
  {\bibinfo {title} {Single-shot realization of nonadiabatic holonomic gates
  with a superconducting xmon qutrit},\ }\href
  {https://doi.org/10.1088/1367-2630/ab2e26} {\bibfield  {journal} {\bibinfo
  {journal} {New Journal of Physics}\ }\textbf {\bibinfo {volume} {21}},\
  \bibinfo {pages} {073024} (\bibinfo {year} {2019})}\BibitemShut {NoStop}%
\bibitem [{\citenamefont {Vepsäläinen}\ \emph {et~al.}(2016)\citenamefont
  {Vepsäläinen}, \citenamefont {Danilin}, \citenamefont {Paladino},
  \citenamefont {Falci},\ and\ \citenamefont {Paraoanu}}]{Veps_l_inen_2016}%
  \BibitemOpen
  \bibfield  {author} {\bibinfo {author} {\bibfnamefont {A.}~\bibnamefont
  {Vepsäläinen}}, \bibinfo {author} {\bibfnamefont {S.}~\bibnamefont
  {Danilin}}, \bibinfo {author} {\bibfnamefont {E.}~\bibnamefont {Paladino}},
  \bibinfo {author} {\bibfnamefont {G.}~\bibnamefont {Falci}},\ and\ \bibinfo
  {author} {\bibfnamefont {G.}~\bibnamefont {Paraoanu}},\ }\bibfield  {title}
  {\bibinfo {title} {Quantum control in qutrit systems using hybrid rabi-stirap
  pulses},\ }\href {https://doi.org/10.3390/photonics3040062} {\bibfield
  {journal} {\bibinfo  {journal} {Photonics}\ }\textbf {\bibinfo {volume}
  {3}},\ \bibinfo {pages} {62} (\bibinfo {year} {2016})}\BibitemShut {NoStop}%
\bibitem [{\citenamefont {Morvan}\ \emph {et~al.}(2020)\citenamefont {Morvan},
  \citenamefont {Ramasesh}, \citenamefont {Blok}, \citenamefont {Kreikebaum},
  \citenamefont {O'Brien}, \citenamefont {Chen}, \citenamefont {Mitchell},
  \citenamefont {Naik}, \citenamefont {Santiago},\ and\ \citenamefont
  {Siddiqi}}]{morvan2020qutrit}%
  \BibitemOpen
  \bibfield  {author} {\bibinfo {author} {\bibfnamefont {A.}~\bibnamefont
  {Morvan}}, \bibinfo {author} {\bibfnamefont {V.~V.}\ \bibnamefont
  {Ramasesh}}, \bibinfo {author} {\bibfnamefont {M.~S.}\ \bibnamefont {Blok}},
  \bibinfo {author} {\bibfnamefont {J.~M.}\ \bibnamefont {Kreikebaum}},
  \bibinfo {author} {\bibfnamefont {K.}~\bibnamefont {O'Brien}}, \bibinfo
  {author} {\bibfnamefont {L.}~\bibnamefont {Chen}}, \bibinfo {author}
  {\bibfnamefont {B.~K.}\ \bibnamefont {Mitchell}}, \bibinfo {author}
  {\bibfnamefont {R.~K.}\ \bibnamefont {Naik}}, \bibinfo {author}
  {\bibfnamefont {D.~I.}\ \bibnamefont {Santiago}},\ and\ \bibinfo {author}
  {\bibfnamefont {I.}~\bibnamefont {Siddiqi}},\ }\href@noop {} {\bibinfo
  {title} {Qutrit randomized benchmarking}} (\bibinfo {year} {2020}),\ \Eprint
  {https://arxiv.org/abs/2008.09134} {arXiv:2008.09134 [quant-ph]} \BibitemShut
  {NoStop}%
\bibitem [{\citenamefont {{Blok}}\ \emph {et~al.}(2020)\citenamefont {{Blok}},
  \citenamefont {{Ramasesh}}, \citenamefont {{Schuster}}, \citenamefont
  {{O'Brien}}, \citenamefont {{Kreikebaum}}, \citenamefont {{Dahlen}},
  \citenamefont {{Morvan}}, \citenamefont {{Yoshida}}, \citenamefont {{Yao}},\
  and\ \citenamefont {{Siddiqi}}}]{2020arXiv200303307B}%
  \BibitemOpen
  \bibfield  {author} {\bibinfo {author} {\bibfnamefont {M.~S.}\ \bibnamefont
  {{Blok}}}, \bibinfo {author} {\bibfnamefont {V.~V.}\ \bibnamefont
  {{Ramasesh}}}, \bibinfo {author} {\bibfnamefont {T.}~\bibnamefont
  {{Schuster}}}, \bibinfo {author} {\bibfnamefont {K.}~\bibnamefont
  {{O'Brien}}}, \bibinfo {author} {\bibfnamefont {J.~M.}\ \bibnamefont
  {{Kreikebaum}}}, \bibinfo {author} {\bibfnamefont {D.}~\bibnamefont
  {{Dahlen}}}, \bibinfo {author} {\bibfnamefont {A.}~\bibnamefont {{Morvan}}},
  \bibinfo {author} {\bibfnamefont {B.}~\bibnamefont {{Yoshida}}}, \bibinfo
  {author} {\bibfnamefont {N.~Y.}\ \bibnamefont {{Yao}}},\ and\ \bibinfo
  {author} {\bibfnamefont {I.}~\bibnamefont {{Siddiqi}}},\ }\bibfield  {title}
  {\bibinfo {title} {{Quantum Information Scrambling in a Superconducting
  Qutrit Processor}},\ }\href@noop {} {\bibfield  {journal} {\bibinfo
  {journal} {arXiv e-prints}\ ,\ \bibinfo {eid} {arXiv:2003.03307}} (\bibinfo
  {year} {2020})},\ \Eprint {https://arxiv.org/abs/2003.03307}
  {arXiv:2003.03307 [quant-ph]} \BibitemShut {NoStop}%
\bibitem [{\citenamefont {Senko}\ \emph {et~al.}(2015)\citenamefont {Senko},
  \citenamefont {Richerme}, \citenamefont {Smith}, \citenamefont {Lee},
  \citenamefont {Cohen}, \citenamefont {Retzker},\ and\ \citenamefont
  {Monroe}}]{PhysRevX.5.021026}%
  \BibitemOpen
  \bibfield  {author} {\bibinfo {author} {\bibfnamefont {C.}~\bibnamefont
  {Senko}}, \bibinfo {author} {\bibfnamefont {P.}~\bibnamefont {Richerme}},
  \bibinfo {author} {\bibfnamefont {J.}~\bibnamefont {Smith}}, \bibinfo
  {author} {\bibfnamefont {A.}~\bibnamefont {Lee}}, \bibinfo {author}
  {\bibfnamefont {I.}~\bibnamefont {Cohen}}, \bibinfo {author} {\bibfnamefont
  {A.}~\bibnamefont {Retzker}},\ and\ \bibinfo {author} {\bibfnamefont
  {C.}~\bibnamefont {Monroe}},\ }\bibfield  {title} {\bibinfo {title}
  {Realization of a quantum integer-spin chain with controllable
  interactions},\ }\href {https://doi.org/10.1103/PhysRevX.5.021026} {\bibfield
   {journal} {\bibinfo  {journal} {Phys. Rev. X}\ }\textbf {\bibinfo {volume}
  {5}},\ \bibinfo {pages} {021026} (\bibinfo {year} {2015})}\BibitemShut
  {NoStop}%
\bibitem [{\citenamefont {Wang}\ \emph {et~al.}(2020)\citenamefont {Wang},
  \citenamefont {Curtis}, \citenamefont {Lester}, \citenamefont {Zhang},
  \citenamefont {Gao}, \citenamefont {Freeze}, \citenamefont {Batista},
  \citenamefont {Vaccaro}, \citenamefont {Chuang}, \citenamefont {Frunzio},
  \citenamefont {Jiang}, \citenamefont {Girvin},\ and\ \citenamefont
  {Schoelkopf}}]{PhysRevX.10.021060}%
  \BibitemOpen
  \bibfield  {author} {\bibinfo {author} {\bibfnamefont {C.~S.}\ \bibnamefont
  {Wang}}, \bibinfo {author} {\bibfnamefont {J.~C.}\ \bibnamefont {Curtis}},
  \bibinfo {author} {\bibfnamefont {B.~J.}\ \bibnamefont {Lester}}, \bibinfo
  {author} {\bibfnamefont {Y.}~\bibnamefont {Zhang}}, \bibinfo {author}
  {\bibfnamefont {Y.~Y.}\ \bibnamefont {Gao}}, \bibinfo {author} {\bibfnamefont
  {J.}~\bibnamefont {Freeze}}, \bibinfo {author} {\bibfnamefont {V.~S.}\
  \bibnamefont {Batista}}, \bibinfo {author} {\bibfnamefont {P.~H.}\
  \bibnamefont {Vaccaro}}, \bibinfo {author} {\bibfnamefont {I.~L.}\
  \bibnamefont {Chuang}}, \bibinfo {author} {\bibfnamefont {L.}~\bibnamefont
  {Frunzio}}, \bibinfo {author} {\bibfnamefont {L.}~\bibnamefont {Jiang}},
  \bibinfo {author} {\bibfnamefont {S.~M.}\ \bibnamefont {Girvin}},\ and\
  \bibinfo {author} {\bibfnamefont {R.~J.}\ \bibnamefont {Schoelkopf}},\
  }\bibfield  {title} {\bibinfo {title} {Efficient multiphoton sampling of
  molecular vibronic spectra on a superconducting bosonic processor},\ }\href
  {https://doi.org/10.1103/PhysRevX.10.021060} {\bibfield  {journal} {\bibinfo
  {journal} {Phys. Rev. X}\ }\textbf {\bibinfo {volume} {10}},\ \bibinfo
  {pages} {021060} (\bibinfo {year} {2020})}\BibitemShut {NoStop}%
\bibitem [{\citenamefont {Zhang}\ \emph {et~al.}(2021)\citenamefont {Zhang},
  \citenamefont {Meurice},\ and\ \citenamefont {Tsai}}]{zhang2021truncation}%
  \BibitemOpen
  \bibfield  {author} {\bibinfo {author} {\bibfnamefont {J.}~\bibnamefont
  {Zhang}}, \bibinfo {author} {\bibfnamefont {Y.}~\bibnamefont {Meurice}},\
  and\ \bibinfo {author} {\bibfnamefont {S.-W.}\ \bibnamefont {Tsai}},\
  }\bibfield  {title} {\bibinfo {title} {Truncation effects in the charge
  representation of the o(2) model},\ }\href@noop {} {\  (\bibinfo {year}
  {2021})},\ \Eprint {https://arxiv.org/abs/2104.06342} {arXiv:2104.06342
  [cond-mat.quant-gas]} \BibitemShut {NoStop}%
\bibitem [{\citenamefont {Creutz}\ \emph {et~al.}(1979)\citenamefont {Creutz},
  \citenamefont {Jacobs},\ and\ \citenamefont {Rebbi}}]{PhysRevD.20.1915}%
  \BibitemOpen
  \bibfield  {author} {\bibinfo {author} {\bibfnamefont {M.}~\bibnamefont
  {Creutz}}, \bibinfo {author} {\bibfnamefont {L.}~\bibnamefont {Jacobs}},\
  and\ \bibinfo {author} {\bibfnamefont {C.}~\bibnamefont {Rebbi}},\ }\bibfield
   {title} {\bibinfo {title} {Monte carlo study of abelian lattice gauge
  theories},\ }\href {https://doi.org/10.1103/PhysRevD.20.1915} {\bibfield
  {journal} {\bibinfo  {journal} {Phys. Rev. D}\ }\textbf {\bibinfo {volume}
  {20}},\ \bibinfo {pages} {1915} (\bibinfo {year} {1979})}\BibitemShut
  {NoStop}%
\bibitem [{\citenamefont {Lloyd}(1996)}]{Lloyd1073}%
  \BibitemOpen
  \bibfield  {author} {\bibinfo {author} {\bibfnamefont {S.}~\bibnamefont
  {Lloyd}},\ }\bibfield  {title} {\bibinfo {title} {Universal quantum
  simulators},\ }\href {https://doi.org/10.1126/science.273.5278.1073}
  {\bibfield  {journal} {\bibinfo  {journal} {Science}\ }\textbf {\bibinfo
  {volume} {273}},\ \bibinfo {pages} {1073} (\bibinfo {year}
  {1996})}\BibitemShut {NoStop}%
\bibitem [{\citenamefont {Tilma}\ and\ \citenamefont
  {Sudarshan}(2002)}]{Tilma_2002}%
  \BibitemOpen
  \bibfield  {author} {\bibinfo {author} {\bibfnamefont {T.}~\bibnamefont
  {Tilma}}\ and\ \bibinfo {author} {\bibfnamefont {E.~C.~G.}\ \bibnamefont
  {Sudarshan}},\ }\bibfield  {title} {\bibinfo {title} {Generalized euler angle
  parametrization forsu(n)},\ }\href
  {https://doi.org/10.1088/0305-4470/35/48/316} {\bibfield  {journal} {\bibinfo
   {journal} {Journal of Physics A: Mathematical and General}\ }\textbf
  {\bibinfo {volume} {35}},\ \bibinfo {pages} {10467–10501} (\bibinfo {year}
  {2002})}\BibitemShut {NoStop}%
\bibitem [{\citenamefont {Di}\ and\ \citenamefont {Wei}(2013)}]{Di_2013}%
  \BibitemOpen
  \bibfield  {author} {\bibinfo {author} {\bibfnamefont {Y.-M.}\ \bibnamefont
  {Di}}\ and\ \bibinfo {author} {\bibfnamefont {H.-R.}\ \bibnamefont {Wei}},\
  }\bibfield  {title} {\bibinfo {title} {Synthesis of multivalued quantum logic
  circuits by elementary gates},\ }\bibfield  {journal} {\bibinfo  {journal}
  {Physical Review A}\ }\textbf {\bibinfo {volume} {87}},\ \href
  {https://doi.org/10.1103/physreva.87.012325} {10.1103/physreva.87.012325}
  (\bibinfo {year} {2013})\BibitemShut {NoStop}%
\bibitem [{\citenamefont {Vatan}\ and\ \citenamefont
  {Williams}(2004)}]{Vatan_2004}%
  \BibitemOpen
  \bibfield  {author} {\bibinfo {author} {\bibfnamefont {F.}~\bibnamefont
  {Vatan}}\ and\ \bibinfo {author} {\bibfnamefont {C.}~\bibnamefont
  {Williams}},\ }\bibfield  {title} {\bibinfo {title} {Optimal quantum circuits
  for general two-qubit gates},\ }\bibfield  {journal} {\bibinfo  {journal}
  {Physical Review A}\ }\textbf {\bibinfo {volume} {69}},\ \href
  {https://doi.org/10.1103/physreva.69.032315} {10.1103/physreva.69.032315}
  (\bibinfo {year} {2004})\BibitemShut {NoStop}%
\bibitem [{\citenamefont {Miller}\ \emph {et~al.}(2018)\citenamefont {Miller},
  \citenamefont {Holz}, \citenamefont {Kampermann},\ and\ \citenamefont
  {Bruß}}]{Miller_2018}%
  \BibitemOpen
  \bibfield  {author} {\bibinfo {author} {\bibfnamefont {D.}~\bibnamefont
  {Miller}}, \bibinfo {author} {\bibfnamefont {T.}~\bibnamefont {Holz}},
  \bibinfo {author} {\bibfnamefont {H.}~\bibnamefont {Kampermann}},\ and\
  \bibinfo {author} {\bibfnamefont {D.}~\bibnamefont {Bruß}},\ }\bibfield
  {title} {\bibinfo {title} {Propagation of generalized pauli errors in qudit
  clifford circuits},\ }\bibfield  {journal} {\bibinfo  {journal} {Physical
  Review A}\ }\textbf {\bibinfo {volume} {98}},\ \href
  {https://doi.org/10.1103/physreva.98.052316} {10.1103/physreva.98.052316}
  (\bibinfo {year} {2018})\BibitemShut {NoStop}%
\bibitem [{\citenamefont {{Checinska}}\ and\ \citenamefont
  {{Wodkiewicz}}(2006)}]{2006quant.ph.10127C}%
  \BibitemOpen
  \bibfield  {author} {\bibinfo {author} {\bibfnamefont {A.}~\bibnamefont
  {{Checinska}}}\ and\ \bibinfo {author} {\bibfnamefont {K.}~\bibnamefont
  {{Wodkiewicz}}},\ }\bibfield  {title} {\bibinfo {title} {{Noisy Qutrit
  Channels}},\ }\href@noop {} {\bibfield  {journal} {\bibinfo  {journal} {arXiv
  e-prints}\ ,\ \bibinfo {eid} {quant-ph/0610127}} (\bibinfo {year} {2006})},\
  \Eprint {https://arxiv.org/abs/quant-ph/0610127} {arXiv:quant-ph/0610127
  [quant-ph]} \BibitemShut {NoStop}%
\bibitem [{\citenamefont {Nielsen}\ and\ \citenamefont
  {Chuang}(2010)}]{nielsen_chuang_2010}%
  \BibitemOpen
  \bibfield  {author} {\bibinfo {author} {\bibfnamefont {M.~A.}\ \bibnamefont
  {Nielsen}}\ and\ \bibinfo {author} {\bibfnamefont {I.~L.}\ \bibnamefont
  {Chuang}},\ }\href {https://doi.org/10.1017/CBO9780511976667} {\emph
  {\bibinfo {title} {Quantum Computation and Quantum Information: 10th
  Anniversary Edition}}}\ (\bibinfo  {publisher} {Cambridge University Press},\
  \bibinfo {year} {2010})\BibitemShut {NoStop}%
\end{thebibliography}%
\end{document}